\documentclass[zpreprint,zbstepj,british]{zeus_K2.9paper}
\usepackage{graphicx}
\usepackage{booktabs}
\usepackage{multirow}
\usepackage{comment}
\usepackage{caption}
\usepackage{subcaption}
\usepackage{lineno}
\usepackage{stackengine}

\usepackage{rotating}
\usepackage{amsmath}
\usepackage{verbatim}
\usepackage{stackengine}
\usepackage{pdfpages}
\usepackage{longtable}
\usepackage{ulem}

\usepackage{zeusdet_hepunits}
\usepackage{array, tabularx, booktabs, longtable}
\def\citeCTD{{\cite{%
nim:a279:290,*npps:b32:181,*nim:a338:254%
}}\xspace}
\def\citeMVD{{\cite{%
nim:a581:656%
}}\xspace}
\def\citeSTT{{\cite{%
nim:a535:191%
}}\xspace}
\def\citeHES{{\cite{%
nim:a277:176%
}}\xspace}
\def\citeSixmT{{\cite{%
thesis:gosau:2007%
}}\xspace}
\def\citeBAC{{\cite{%
nim:a300:480%
}}\xspace}
\def\citePCAL{{\cite{%
desy-92-066,*zfp:c63:391,*acpp:b32:2025%
}}\xspace}
\def\citeSPECTRO{{\cite{%
nim:a565:572%
}}\xspace}

\newcommand{\pp}{\ensuremath{pp}\xspace}
\newcommand{\Nrec}{\ensuremath{N_{\textrm{rec}}}\xspace}
\newcommand{\Nch}{\ensuremath{N_{\textrm{ch}}}\xspace}
\newcommand{\cn}[1]{\ensuremath{c_{#1}{\scriptstyle \{2\}}}\xspace}
\newcommand{\Cdetadphi}{\ensuremath{C(\Delta\eta, \Delta\varphi)}\xspace}
\newcommand{\deta}{\ensuremath{|\Delta \eta|}\xspace}

\newcommand{\pT}{\ensuremath{p_{\rm T}}\xspace}
\newcommand{\meanpt}{\ensuremath{\left< \pT \right>}\xspace}
\newcommand{\ep}{\ensuremath{ep}\xspace}

\newcommand{\wdphi}{\ensuremath{w_{\Delta\varphi}}\xspace}

\newcommand{\dsyst}{\ensuremath{\delta_{\textrm{syst}}}\xspace}
\newcommand{\dnc}{\ensuremath{\delta_{\textrm{nc}}}\xspace}
\newcommand{\dncAri}{\ensuremath{\delta_{\textrm{nc}}^{\textrm{ARIADNE}}}\xspace}
\newcommand{\dncLepto}{\ensuremath{\delta_{\textrm{nc}}^{\textrm{LEPTO}}}\xspace}

\usepackage{array}

\usepackage{dcolumn}
\usepackage{color,soul} %to highlight text
\usepackage{color}
\newcolumntype{d}[1]{D{.}{.}{#1} }%start tables 
\begin{document}
%------------------------------------------------------------------------------
%       Title sheet
%------------------------------------------------------------------------------
%
%
%
\prepnum{DESY--19--174}
\draftversion{}
\prepdate{March 2020}%\today                     %MMMM YYYY

\title{\Large Two-particle azimuthal correlations as a probe of collective behaviour in deep inelastic \textit{ep} scattering at HERA}
                  
\author{ZEUS Collaboration}
\date{}              % The line is needed.  Add parameter \today for date.

\maketitle

%\linenumbers

%
% If you use the package units instead of hepunits you have to enclose
% the values in square brackets and change \invpb to \pbi and \GeV to \Gev
% e.g. \unit[47.7]{\pbi}
% If the quantity is not in math mode and the unit contains math mode
% characters such as superscripts it must be contained in $...$
%
\begin{abstract}\noindent
{Two-particle azimuthal correlations have been measured in neutral current deep \mbox{inelastic} \ep scattering with virtuality $Q^2 > 5\GeV^2$ at a centre-of-mass energy \mbox{$\sqrt{s}=318\GeV$} recorded with the ZEUS detector at HERA.
The correlations of charged particles have been measured in the range of laboratory pseudorapidity $-1.5 < \eta < 2.0$ and transverse momentum $0.1 < \pT~< 5.0\GeV$~and event multiplicities \Nch up to six times larger than the average $\left< \Nch \right>\approx5$.
The two-particle correlations have been measured in terms of the angular observables $\cn{n}=\left< \left< \cos{ n \Delta\varphi} \right> \right>$, where $n$ is between 1 and 4 and $\Delta\varphi$ is the relative azimuthal angle between the two particles.
Comparisons with available models of deep inelastic scattering, which are tuned to reproduce inclusive particle production, suggest that the measured two-particle correlations are dominated by contributions from multijet production.
The correlations observed here do not indicate the kind of collective behaviour recently observed at the highest RHIC and LHC energies in high-multiplicity hadronic collisions.
}

\end{abstract}

\thispagestyle{empty}
%------------------------------------------------------------------------------
%       Authors - you may have to play with \clearpage and \cleardoublepage 
%       in order to get the main text to start on the correct page
%------------------------------------------------------------------------------
\clearpage

%===================================================================
%
%  MEMBER NAME  AUTH183 (ZEUS)     M  TEX
%
%  JH.: transformed to a format, which is suited as input for
%       CONVERT, which automatically creates author-indices
%
%  Don't remove lines starting with a percent sign %,
%  CONVERT may need them urgently !
%  
%=====================================================================

                                                   %
\begin{center}
{                      \Large  The ZEUS Collaboration              }
\end{center}

{\small\raggedright

%  members:

I.~Abt$^{20}$, 
L.~Adamczyk$^{7}$, 
R. Aggarwal$^{3, b}$, 
V.~Aushev$^{18}$, 
O.~Behnke$^{9}$, 
U.~Behrens$^{9}$, 
A.~Bertolin$^{22}$, 
I.~Bloch$^{10}$, 
I.~Brock$^{2}$, 
N.H.~Brook$^{29, m}$, 
R.~Brugnera$^{23}$, 
A.~Bruni$^{1}$, 
P.J.~Bussey$^{11}$, 
A.~Caldwell$^{20}$, 
M.~Capua$^{4}$, 
C.D.~Catterall$^{33}$, 
J.~Chwastowski$^{6}$, 
J.~Ciborowski$^{30, o}$, 
R.~Ciesielski$^{9, d}$, 
A.M.~Cooper-Sarkar$^{21}$, 
M.~Corradi$^{1, a}$, 
R.K.~Dementiev$^{19}$, 
S.~Dusini$^{22}$, 
J.~Ferrando$^{9}$, 
S.~Floerchinger$^{15}$,
B.~Foster$^{21, j}$, 
E.~Gallo$^{13, k}$, 
D.~Gangadharan$^{14}$, 
A.~Garfagnini$^{23}$, 
A.~Geiser$^{9}$, 
L.K.~Gladilin$^{19}$, 
Yu.A.~Golubkov$^{19}$, 
G.~Grzelak$^{30}$, 
C.~Gwenlan$^{21}$, 
D.~Hochman$^{32}$, 
N.Z.~Jomhari$^{9}$, 
I.~Kadenko$^{18}$, 
S.~Kananov$^{24}$, 
U.~Karshon$^{32}$, 
P.~Kaur$^{3, c}$, 
R.~Klanner$^{13}$, 
U.~Klein$^{9, e}$, 
I.A.~Korzhavina$^{19}$, 
N.~Kovalchuk$^{13}$, 
H.~Kowalski$^{9}$, 
O.~Kuprash$^{9, f}$, 
M.~Kuze$^{26}$, 
B.B.~Levchenko$^{19}$, 
A.~Levy$^{24}$, 
B.~L\"ohr$^{9}$, 
E.~Lohrmann$^{13}$, 
A.~Longhin$^{23}$, 
O.Yu.~Lukina$^{19}$, 
I.~Makarenko$^{9}$, 
J.~Malka$^{9, g}$, 
S.~Masciocchi$^{12, i}$, 
K.~Nagano$^{16}$, 
J.D.~Nam$^{25}$, 
J.~Onderwaater$^{14, l}$, 
Yu.~Onishchuk$^{18}$, 
E.~Paul$^{2}$, 
I.~Pidhurskyi$^{18}$, 
A.~Polini$^{1}$, 
M.~Przybycie\'n$^{7}$, 
A.~Quintero$^{25}$, 
M.~Ruspa$^{28}$, 
D.H.~Saxon$^{11}$, 
U.~Schneekloth$^{9}$, 
T.~Sch\"orner-Sadenius$^{9}$, 
I.~Selyuzhenkov$^{12}$, 
M.~Shchedrolosiev$^{18}$, 
L.M.~Shcheglova$^{19}$, 
I.O.~Skillicorn$^{11}$, 
W.~S{\l}omi\'nski$^{8}$, 
A.~Solano$^{27}$, 
L.~Stanco$^{22}$, 
N.~Stefaniuk$^{9}$, 
P.~Stopa$^{6}$, 
B.~Surrow$^{25}$, 
J.~Sztuk-Dambietz$^{13, g}$, 
E.~Tassi$^{4}$, 
K.~Tokushuku$^{16}$, 
M.~Turcato$^{13, g}$, 
O.~Turkot$^{9}$, 
T.~Tymieniecka$^{31}$, 
A.~Verbytskyi$^{20}$, 
W.A.T.~Wan Abdullah$^{5}$, 
K.~Wichmann$^{9}$, 
M.~Wing$^{29, n}$, 
S.~Yamada$^{16}$, 
Y.~Yamazaki$^{17}$, 
A.F.~\.Zarnecki$^{30}$, 
L.~Zawiejski$^{6}$, 
O.~Zenaiev$^{9, h}$ 
\newpage

%       institutes:

{\setlength{\parskip}{0.4em}
\makebox[3ex]{$^{1}$}
\begin{minipage}[t]{14cm}
{\it INFN Bologna, Bologna, Italy}~$^{A}$

\end{minipage}

\makebox[3ex]{$^{2}$}
\begin{minipage}[t]{14cm}
{\it Physikalisches Institut der Universit\"at Bonn,
Bonn, Germany}~$^{B}$

\end{minipage}

\makebox[3ex]{$^{3}$}
\begin{minipage}[t]{14cm}
{\it Panjab University, Department of Physics, Chandigarh, India}

\end{minipage}

\makebox[3ex]{$^{4}$}
\begin{minipage}[t]{14cm}
{\it Calabria University,
Physics Department and INFN, Cosenza, Italy}~$^{A}$

\end{minipage}

\makebox[3ex]{$^{5}$}
\begin{minipage}[t]{14cm}
{\it National Centre for Particle Physics, Universiti Malaya, 50603 Kuala Lumpur, Malaysia}~$^{C}$

\end{minipage}

\makebox[3ex]{$^{6}$}
\begin{minipage}[t]{14cm}
{\it The Henryk Niewodniczanski Institute of Nuclear Physics, Polish Academy of \\
Sciences, Krakow, Poland}

\end{minipage}

\makebox[3ex]{$^{7}$}
\begin{minipage}[t]{14cm}
{\it AGH University of Science and Technology, Faculty of Physics and Applied Computer
Science, Krakow, Poland}

\end{minipage}

\makebox[3ex]{$^{8}$}
\begin{minipage}[t]{14cm}
{\it Department of Physics, Jagellonian University, Krakow, Poland}~$^{D}$

\end{minipage}

\makebox[3ex]{$^{9}$}
\begin{minipage}[t]{14cm}
{\it Deutsches Elektronen-Synchrotron DESY, Hamburg, Germany}

\end{minipage}

\makebox[3ex]{$^{10}$}
\begin{minipage}[t]{14cm}
{\it Deutsches Elektronen-Synchrotron DESY, Zeuthen, Germany}

\end{minipage}

\makebox[3ex]{$^{11}$}
\begin{minipage}[t]{14cm}
{\it School of Physics and Astronomy, University of Glasgow,
Glasgow, United Kingdom}~$^{E}$

\end{minipage}

\makebox[3ex]{$^{12}$}
\begin{minipage}[t]{14cm}
{\it GSI Helmholtzzentrum f\"{u}r Schwerionenforschung GmbH, Darmstadt, Germany}~$^{K}$

\end{minipage}

\makebox[3ex]{$^{13}$}
\begin{minipage}[t]{14cm}
{\it Hamburg University, Institute of Experimental Physics, Hamburg,
Germany}~$^{F}$

\end{minipage}

\makebox[3ex]{$^{14}$}
\begin{minipage}[t]{14cm}
{\it Physikalisches Institute, University of Heidelberg, Heidelberg, Germany~$^{K}$}

\end{minipage}

\makebox[3ex]{$^{15}$}
\begin{minipage}[t]{14cm}
{\it Institute for Theoretical Physics, University of Heidelberg, Heidelberg, Germany~$^{K}$}

\end{minipage}

\makebox[3ex]{$^{16}$}
\begin{minipage}[t]{14cm}
{\it Institute of Particle and Nuclear Studies, KEK,
Tsukuba, Japan}~$^{G}$

\end{minipage}

\makebox[3ex]{$^{17}$}
\begin{minipage}[t]{14cm}
{\it Department of Physics, Kobe University, Kobe, Japan}~$^{G}$

\end{minipage}

\makebox[3ex]{$^{18}$}
\begin{minipage}[t]{14cm}
{\it Department of Nuclear Physics, National Taras Shevchenko University of Kyiv, Kyiv, Ukraine}

\end{minipage}

\makebox[3ex]{$^{19}$}
\begin{minipage}[t]{14cm}
{\it Lomonosov Moscow State University, Skobeltsyn Institute of Nuclear Physics,
Moscow, Russia}

\end{minipage}

\makebox[3ex]{$^{20}$}
\begin{minipage}[t]{14cm}
{\it Max-Planck-Institut f\"ur Physik, M\"unchen, Germany}

\end{minipage}

\makebox[3ex]{$^{21}$}
\begin{minipage}[t]{14cm}
{\it Department of Physics, University of Oxford,
Oxford, United Kingdom}~$^{E}$

\end{minipage}

\makebox[3ex]{$^{22}$}
\begin{minipage}[t]{14cm}
{\it INFN Padova, Padova, Italy}~$^{A}$

\end{minipage}

\makebox[3ex]{$^{23}$}
\begin{minipage}[t]{14cm}
{\it Dipartimento di Fisica e Astronomia dell' Universit\`a and INFN,
Padova, Italy}~$^{A}$

\end{minipage}

\makebox[3ex]{$^{24}$}
\begin{minipage}[t]{14cm}
{\it Raymond and Beverly Sackler Faculty of Exact Sciences, School of Physics, \\
Tel Aviv University, Tel Aviv, Israel}~$^{H}$

\end{minipage}

\makebox[3ex]{$^{25}$}
\begin{minipage}[t]{14cm}
{\it Department of Physics, Temple University, Philadelphia, PA 19122, USA}~$^{I}$

\end{minipage}

\makebox[3ex]{$^{26}$}
\begin{minipage}[t]{14cm}
{\it Department of Physics, Tokyo Institute of Technology,
Tokyo, Japan}~$^{G}$

\end{minipage}

\makebox[3ex]{$^{27}$}
\begin{minipage}[t]{14cm}
{\it Universit\`a di Torino and INFN, Torino, Italy}~$^{A}$

\end{minipage}

\makebox[3ex]{$^{28}$}
\begin{minipage}[t]{14cm}
{\it Universit\`a del Piemonte Orientale, Novara, and INFN, Torino,
Italy}~$^{A}$

\end{minipage}

\makebox[3ex]{$^{29}$}
\begin{minipage}[t]{14cm}
{\it Physics and Astronomy Department, University College London,
London, United Kingdom}~$^{E}$

\end{minipage}

\makebox[3ex]{$^{30}$}
\begin{minipage}[t]{14cm}
{\it Faculty of Physics, University of Warsaw, Warsaw, Poland}

\end{minipage}

\makebox[3ex]{$^{31}$}
\begin{minipage}[t]{14cm}
{\it National Centre for Nuclear Research, Warsaw, Poland}

\end{minipage}

\makebox[3ex]{$^{32}$}
\begin{minipage}[t]{14cm}
{\it Department of Particle Physics and Astrophysics, Weizmann
Institute, Rehovot, Israel}

\end{minipage}

\makebox[3ex]{$^{33}$}
\begin{minipage}[t]{14cm}
{\it Department of Physics, York University, Ontario, Canada M3J 1P3}~$^{J}$

\end{minipage}

}
\vspace{3em}

%  references concerning institutes;

{\setlength{\parskip}{0.4em}\raggedright
\makebox[3ex]{$^{ A}$}
\begin{minipage}[t]{14cm}
 supported by the Italian National Institute for Nuclear Physics (INFN) \
\end{minipage}

\makebox[3ex]{$^{ B}$}
\begin{minipage}[t]{14cm}
 supported by the German Federal Ministry for Education and Research (BMBF), under
 contract No.\ 05 H09PDF\
\end{minipage}

\makebox[3ex]{$^{ C}$}
\begin{minipage}[t]{14cm}
 supported by HIR grant UM.C/625/1/HIR/149 and UMRG grants RU006-2013, RP012A-13AFR and RP012B-13AFR from
 Universiti Malaya, and ERGS grant ER004-2012A from the Ministry of Education, Malaysia\
\end{minipage}

\makebox[3ex]{$^{ D}$}
\begin{minipage}[t]{14cm}
supported by the Polish National Science Centre (NCN) grant no.\ DEC-2014/13/B/ST2/02486
\end{minipage}

\makebox[3ex]{$^{ E}$}
\begin{minipage}[t]{14cm}
 supported by the Science and Technology Facilities Council, UK\
\end{minipage}

\makebox[3ex]{$^{ F}$}
\begin{minipage}[t]{14cm}
 supported by the German Federal Ministry for Education and Research (BMBF), under
 contract No.\ 05h09GUF, and the SFB 676 of the Deutsche Forschungsgemeinschaft (DFG) \
\end{minipage}

\makebox[3ex]{$^{ G}$}
\begin{minipage}[t]{14cm}
 supported by the Japanese Ministry of Education, Culture, Sports, Science and Technology
 (MEXT) and its grants for Scientific Research\
\end{minipage}

\makebox[3ex]{$^{ H}$}
\begin{minipage}[t]{14cm}
 supported by the Israel Science Foundation\
\end{minipage}

\makebox[3ex]{$^{ I}$}
\begin{minipage}[t]{14cm}
 supported in part by the Office of Nuclear Physics within the U.S.\ DOE Office of Science
\end{minipage}

\makebox[3ex]{$^{ J}$}
\begin{minipage}[t]{14cm}
 supported by the Natural Sciences and Engineering Research Council of Canada (NSERC) \
\end{minipage}

\makebox[3ex]{$^{ K}$}
\begin{minipage}[t]{14cm}
this work is part of and supported by the DFG Collaborative Research 
Centre ``SFB 1225 (ISOQUANT)''\
\end{minipage}

}

\pagebreak[4]
{\setlength{\parskip}{0.4em}

%  references concerning members;

\makebox[3ex]{$^{ a}$}
\begin{minipage}[t]{14cm}
now at INFN Roma, Italy\
\end{minipage}

\makebox[3ex]{$^{ b}$}
\begin{minipage}[t]{14cm}
now at DST-Inspire Faculty, Department of Technology, SPPU, India\
\end{minipage}

\makebox[3ex]{$^{ c}$}
\begin{minipage}[t]{14cm}
now at Sant Longowal Institute of Engineering and Technology, Longowal, Punjab, India\
\end{minipage}

\makebox[3ex]{$^{ d}$}
\begin{minipage}[t]{14cm}
now at Rockefeller University, New York, NY 10065, USA\
\end{minipage}

\makebox[3ex]{$^{ e}$}
\begin{minipage}[t]{14cm}
now at University of Liverpool, United Kingdom\
\end{minipage}

\makebox[3ex]{$^{ f}$}
\begin{minipage}[t]{14cm}
now at University of Freiburg, Freiburg, Germany\
\end{minipage}

\makebox[3ex]{$^{ g}$}
\begin{minipage}[t]{14cm}
now at European X-ray Free-Electron Laser facility GmbH, Hamburg, Germany\
\end{minipage}

\makebox[3ex]{$^{ h}$}
\begin{minipage}[t]{14cm}
now at Hamburg University, Hamburg, Germany\
\end{minipage}

\makebox[3ex]{$^{ i}$}
\begin{minipage}[t]{14cm}
also at Physikalisches Institute of the University of Heidelberg, Heidelberg,  Germany\
\end{minipage}

\makebox[3ex]{$^{ j}$}
\begin{minipage}[t]{14cm}
also at DESY and University of Hamburg\
\end{minipage}

\makebox[3ex]{$^{ k}$}
\begin{minipage}[t]{14cm}
also at DESY\
\end{minipage}

\makebox[3ex]{$^{ l}$}
\begin{minipage}[t]{14cm}
also at GSI Helmholtzzentrum f\"{u}r Schwerionenforschung GmbH, Darmstadt, Germany\
\end{minipage}

\makebox[3ex]{$^{ m}$}
\begin{minipage}[t]{14cm}
now at University of Bath, United Kingdom\
\end{minipage}

\makebox[3ex]{$^{ n}$}
\begin{minipage}[t]{14cm}
also supported by DESY\
\end{minipage}

\makebox[3ex]{$^{ o}$}
\begin{minipage}[t]{14cm}
also at Lodz University, Poland\
\end{minipage}

}

}

\clearpage
\pagenumbering{arabic}

\section{Introduction}
\label{sec-int}

The search for a new state of matter, the quark--gluon plasma (QGP), has been a major component of the heavy-ion physics programme at many laboratories.
The evidence for its observation in heavy-ion collisions~\cite{Afanasiev:2002mx,Arsene:2004fa,Back:2004je,Adams:2005dq,Adcox:2004mh,Aamodt:2010pa} is strong; one of the key observations being the collective behaviour of final-state particles.
Similar behaviour has recently been observed in high-multiplicity $p+A$ and $pp$ collisions.
This has motivated the first search for such behaviour in \ep collisions at the Hadron Electron Ring Accelerator (HERA).   

The evolution of a QGP in space and time can be described within the framework of relativistic fluid dynamics~\cite{Shuryak:1978ij, Bjorken:1982qr, Heinz:2013th}, employing the thermodynamic and transport properties of quantum chromodynamics (QCD) matter.
The correlated production of the final-state particles reflects this evolution and are referred to as collective behaviour or collectivity. 
Recent measurements~\cite{Abelev:2013vea, Khachatryan:2010gv, Abelev:2012ola, Aad:2012gla, Aad:2015gqa, Adare:2013piz, Adare:2015ctn} at the Relativistic Heavy Ion Collider (RHIC) and the Large Hadron Collider (LHC) have revealed similar collective behaviour in lighter colliding systems at high multiplicity, such as proton nucleus ($p+A$) and even \pp, compared to heavy-ion systems. 
Experimental investigations of the space--time evolution and fragmentation of a multi-parton state formed in \ep collisions at HERA are important to study the presence or absence of collective effects for even smaller interaction regions than those that characterise \pp interactions.

At present it is unclear whether the collectivity observed in different colliding systems is of the same fluid-dynamic origin and how small the interaction region can be until such a description of soft QCD  breaks down.
Fluid-dynamic calculations applied to $A+A$ and high-multiplicity \pp and $p+A$ collisions are able to reproduce reasonably the measurements \cite{Weller:2017tsr,Bozek:2011if,Bozek:2012gr,Bozek:2013uha}. 
This suggests that even in relatively small systems, a state of matter in local thermal equilibrium may be produced, indicating universality of a fluid description.
On the other hand, purely initial-state effects arising from gluon saturation in the colour-glass-condensate framework are also able to describe the qualitative features of the data \cite{Dumitru:2010iy}.

Collisions between fully overlapping heavy nuclei at RHIC and LHC are capable of producing large interaction regions which are of the order of $7$~fm in size.
In \pp and $p+A$ the interaction region is of the order of the proton size, which ranges from its size of around 1 fm, when undisturbed, to a few fm~\cite{Gribov:1968jf, Kharzeev:2017uym}.
The average size of the interaction region in \ep scattering depends on $1/Q$, where the exchanged photon virtuality is defined by the four-momentum difference between the incoming and scattered electron, $Q^2 = -(k - k')^2$.

The terms low and high $Q^2$ are used to distinguish two regimes of particle production in \ep collisions: photoproduction and deep inelastic scattering (DIS)~\cite{Cooper:2003}.
The latter can be further classified into neutral and charged current DIS.
Neutral current (NC) DIS is characterised by the exchange of a virtual photon or $Z$ boson between the incoming electron and proton.
Charged current DIS occurs less frequently, when a $W$ boson is exchanged and a scattered neutrino instead of an electron appears in the final state.
In photoproduction, the electron emits a quasi-real photon ($Q^2 \lesssim \Lambda_{\textrm{QCD}}^2 \approx (200~ \MeV)^2$), which can fluctuate into an extended hadronic object with a size of the order of $1/\Lambda_{\textrm{QCD}}\approx 1$ fm. 
On the other hand, DIS is characterised by the exchange of a highly virtual and more point-like photon ($Q^2 \gg \Lambda_{\textrm{QCD}}^2$) which strikes a single parton with a finer resolution of $1/Q \ll 1$ fm.
Thus, NC DIS provides a unique opportunity to investigate the dynamics of a many-body QCD system produced in smaller interaction regions than those available at RHIC and LHC.
This is complementary to a corresponding investigation using ALEPH data on $e^{+}e^{-}$ collisions at the $Z$ pole~\cite{Badea:2019vey}.

In addition to the size of the interaction region, its spatial anisotropy also plays an important role in the system's space--time evolution.
Depending on the interaction rate during the collective expansion, the spatial anisotropy can be converted into a momentum asymmetry of the produced particles.
In a fluid picture, this arises essentially because pressure gradients accelerate the fluid.
This final-state asymmetry can be quantified with two-particle azimuthal correlations~\cite{Wang:1991qh, Voloshin:1994mz, Borghini:2001vi, Bilandzic:2013kga}, which coincide with the two-particle cumulants:
\begin{equation}
  \cn{n} = \left< \left<
  \cos{ n \Delta\varphi } \right> \right>,
  \label{eq:cn2}
\end{equation}
where $\Delta\varphi = \varphi_1 - \varphi_2$ is the difference in the azimuthal angles of particles 1 and 2.
The inner angular brackets denote an average over all pairs in a given event while the outer angular brackets denote an average over all events.

In the case of collective fluid-like expansion, measurements of the first four harmonics ($n=1-4$) are sensitive to directed, elliptic, triangular, and quadrangular spatial anisotropies, respectively (see review \cite{Voloshin:2008dg} and references therein).
A prominent feature of the collision between partially overlapping heavy ions is the elliptical shape of the interaction region.
This results in the dominance of an elliptical asymmetry, \cn{2}, in the momentum space.
In \ep DIS, the most prominent feature is the recoil of the hadronic system against the electron (momentum conservation), which is reflected in the directed cumulant \cn{1}.

In this paper, measurements of \cn{n} are presented for the first four harmonics as a function of the number of charged particles, the laboratory pseudorapidity difference $\deta=|\eta_1 - \eta_2|$, and the mean transverse momentum $\meanpt=(p_{\rm T,1}+p_{\rm T,2})/2$.

\section{Experimental set-up}
\label{Sec:Experiment}

The NC DIS data sample used in this analysis was taken with the ZEUS detector at HERA during 2003--2007 (HERA II).
During this period, the HERA accelerator collided $27.5\GeV$\ electron/positron\footnote{Hereafter, ``electron'' refers to both electrons and positrons unless otherwise stated. 
HERA operated with electron beams during 2005 and part of 2006, while positrons were accelerated in the other years of this data sample.} beams with $920\GeV$\ proton beams, which yields a nominal centre-of-mass energy of $\sqrt{s} = 318\GeV$.
The integrated luminosity recorded by ZEUS in HERA II at this energy is $366\pm7\,\mathrm{pb}^{-1}$.
% preamble of detector description
\Zdetdesc

% CTD and MVD description
% use argument {\footnote{\ZcoosysBA}} to insert ZEUS coord system definition
\Zctdmvddesc{\footnote{\ZcoosysBA \ZpsrapB}}

% CAL description
\Zcaldesc

% Luminosity description
\Zlumidesc

\section{Event and track selection}
\label{Sec:Selection}

\subsection{Event selection}
\label{Sec:EventSelection}

The ZEUS experiment operated a three-level trigger system~\cite{uproc:chep:1992:222, nim:a580:1257}.
For this analysis, events were selected at the first level if they had an energy deposit
in the CAL consistent with an isolated scattered electron.
At the second level,
a requirement on the energy and longitudinal momentum of the event was
used to select NC DIS event candidates.
At the third level, the full event was
reconstructed and tighter requirements for a DIS electron were made.

NC DIS events are characterised by the observation of a high-energy scattered electron in the CAL and were selected with the following criteria:
\begin{itemize}
\item
the scattered electron was identified using information from the distribution of energy deposited in the CAL, including information from a silicon-detector system embedded in the RCAL and from its associated track, when available.
A neural-network algorithm \cite{nim:a365:508,nim:a391:360} assigned a probability that a given electron candidate was correctly identified as an electron.
The probability was required to be larger than $90\%$;
\item
the event vertex was obtained from a fit to the measured tracks.
To ensure reliable tracking within the CTD, the position of the event vertex along the $Z$ axis, $V_Z$, was required to be within 30 cm of its nominal value.  
The transverse distance of the event vertex from the interaction point was required to be within 0.5 cm.
For the measurements of two-particle correlations, events were required to have at least one track associated with the primary vertex.
The fraction of primary-vertex tracks to the total number of reconstructed tracks was required to be larger than 0.1 to reject beam-gas background;
\item 
the scattered-electron energy, $E_e$, as measured in the CAL, was larger than 10\GeV~to ensure good electron identification;
\item
the virtuality, $Q^2$, as determined by the electron method \cite{proc:hera:1991:23}, was required to be larger than 5 $\GeV^2$, just above the minimum reconstructable value;
\item 
the polar angle of the scattered electron, $\theta_e$, was required to be larger than 1~radian to ensure a reliable measurement in the RCAL or BCAL, which results in an effective upper limit of $Q^2 < 10^4\GeV^2$;
\item
the radial position of the electron on entering the RCAL was required to be larger than 15 cm ($\theta_e \lesssim 3$ radians) to help reject photoproduction events and to ensure a well understood acceptance.
Entrance locations ($X,Y$) with poor acceptance were excluded from the analysis: $5 < X < 11\,$ cm for $Y>0\,$ cm and $-15 < X < -9\,$ cm for $Y<0\,$ cm.
Additionally, the region $-10 < X < 10$ cm for $Y>110\,$ cm contains a significantly higher material budget and was therefore excluded;
\item
for further rejection of photoproduction events, as well as rejection of DIS events with large initial-state photon radiation, a cut based on the quantity $E-p_Z = \sum_i E_i(1 - \cos{\theta_i})$ was applied.
The sum is over all energy-flow objects \cite{epj:c1:81, epj:c6:43} which are formed from calorimeter-cell clusters and tracks, with energy $E_i$ and polar angle $\theta_i$.
For a fully contained NC DIS event, $E-p_Z$ should be twice the beam-electron energy (55\GeV) owing to energy and longitudinal momentum conservation.
This cut also removes background events caused by collisions of protons with residual gas in the beam pipe or the beam pipe itself.
Events were accepted in the interval $47 < E-p_Z < 69\GeV$.
\end{itemize}

These constraints on the scattered electron implicitly remove events with an inelasticity \cite{Cooper:2003} $y = p \, (k-k') / (p \, k) \gtrsim 0.65$, where $p$ represents the four-momentum of the incoming proton.
A total of 45 million NC DIS events were selected for the analysis.
The contamination from photoproduction events has been estimated to be on the order of 1$\%$ as determined from studies of photoproduction Monte Carlo data as well as from events with scattered-electron candidates with the incorrect charge.

\subsection{Track selection}

Reconstructed tracks were used in this analysis if their momentum transverse to the beam-axis and laboratory pseudorapidity were within $0.1 < \pT~< 5.0\GeV$~and $-1.5 < \eta < 2.0$, respectively.
To reject unwanted secondary tracks and false reconstructions, each track was required to have at least one MVD hit.
The track associated to the scattered electron candidate used to identify the NC DIS event was rejected in the correlation analysis.
Owing to occasional showering of the electron in the beam pipe and tracker material, this track is not always uniquely identified. 
Thus, all tracks around the scattered electron candidate within a cone of 0.4 in pseudorapidity and azimuthal angle were rejected.

Tracks corresponding closely to primary charged particles were selected in the analysis by requiring the distances of closest approach to the primary vertex in the transverse (DCA$_{XY}$) and longitudinal (DCA$_{Z}$) directions to be less than 2 cm.
Some secondary tracks, e.g.\ from small-angle scattering in the beam pipe, were therefore retained.
These tracks inherit information about the properties of the corresponding charged primary particles and serve as their substitutes, thereby retaining correlations with other primary particles.

\section{Monte Carlo generators}
\label{Sec:MC}

The LEPTO 6.5 \cite{cpc:101:108} and ARIADNE 4.12 \cite{cpc:71:15} Monte Carlo event generators were used for the comparison of the measurements to known physics mechanisms and for the extraction of efficiency corrections and the associated systematic uncertainties in the correlation analysis.

Both models are interfaced with PYTHIA 5.724 and JETSET 7.410
\cite{cpc:82:74} to handle hadronisation and decays.
The initial hard scattering in ARIADNE is treated with PYTHIA and JETSET.
The LEPTO and ARIADNE generators differ chiefly in the treatment of the QCD cascade process.
In LEPTO, the cascade is treated as a Dokshitzer--Gribov--Lipatov--Altarelli--Parisi (DGLAP)-based backward-evolution shower~\cite{sovjnp:15:438,jetp:46:641,np:b126:298}.  
The ARIADNE generator treats the cascade within the colour dipole model (CDM) \cite{Gustafson:1986db}.
Initial-state radiation (before the central hard scatter) and final-state radiation are treated independently in LEPTO while ARIADNE includes initial- and final-state interference effects.
In the CDM prescription, the production amplitudes of soft gluon emissions are summed coherently while in LEPTO the angular-ordering technique \cite{Mueller:1981ex, Ermolaev:1981cm} is used to emulate the coherence effect.

The selected ZEUS data sample includes a diffractive component \cite{pl:b315:481} where the \ep scattering is mediated by an object carrying the quantum numbers of the vacuum---a Pomeron.
A distinguishing feature of diffractive events is the absence of hadronic activity in the forward direction.
The pseudorapidity of the most-forward energy deposit in the CAL greater than 400 MeV is defined as $\eta_{\textrm{max}}$.
A diffractive and a non-diffractive ARIADNE sample were combined in this analysis using a weighting scheme chosen to reproduce the $\eta_{\textrm{max}}$ distribution in ZEUS data.
The diffractive component was generated with SATRAP~\cite{pr:d59:014017} which was interfaced with ARIADNE and RAPGAP~\cite{cpc:86:147}.  
Purely non-diffractive ARIADNE predictions were also used to illustrate the expected effect of the diffractive component.
The LEPTO Monte Carlo sample only contains a non-diffractive component since a simulation of the diffractive component was not available.

Generated events were passed through GEANT3.21~\cite{tech:cern-dd-ee-84-1} to simulate the ZEUS detector.
Additionally, a fraction of the low-\pT tracks was rejected to compensate for the imperfectly simulated loss of such tracks~\cite{Bachynska:2012,Libov:2013}.
The selection of Monte Carlo events to compute reconstructed distributions and efficiency corrections followed the same criteria as for the reconstructed ZEUS data, see Section~\ref{Sec:EventSelection}.
Primary generated particles were defined as charged hadrons with a mean proper lifetime, $\tau>1$~cm, which were produced directly or from the decay of a particle with $\tau < 1$~cm.
This definition is the same as that used by ALICE at the LHC~\cite{ALICE-PUBLIC-2017-005}.

\section{Comparison of reconstructed data and Monte Carlo}
\label{Sec:RawData}

\begin{figure}[ht]
  \centering
\begin{picture}(450,400)
\put(0,0){\includegraphics[width=\textwidth]{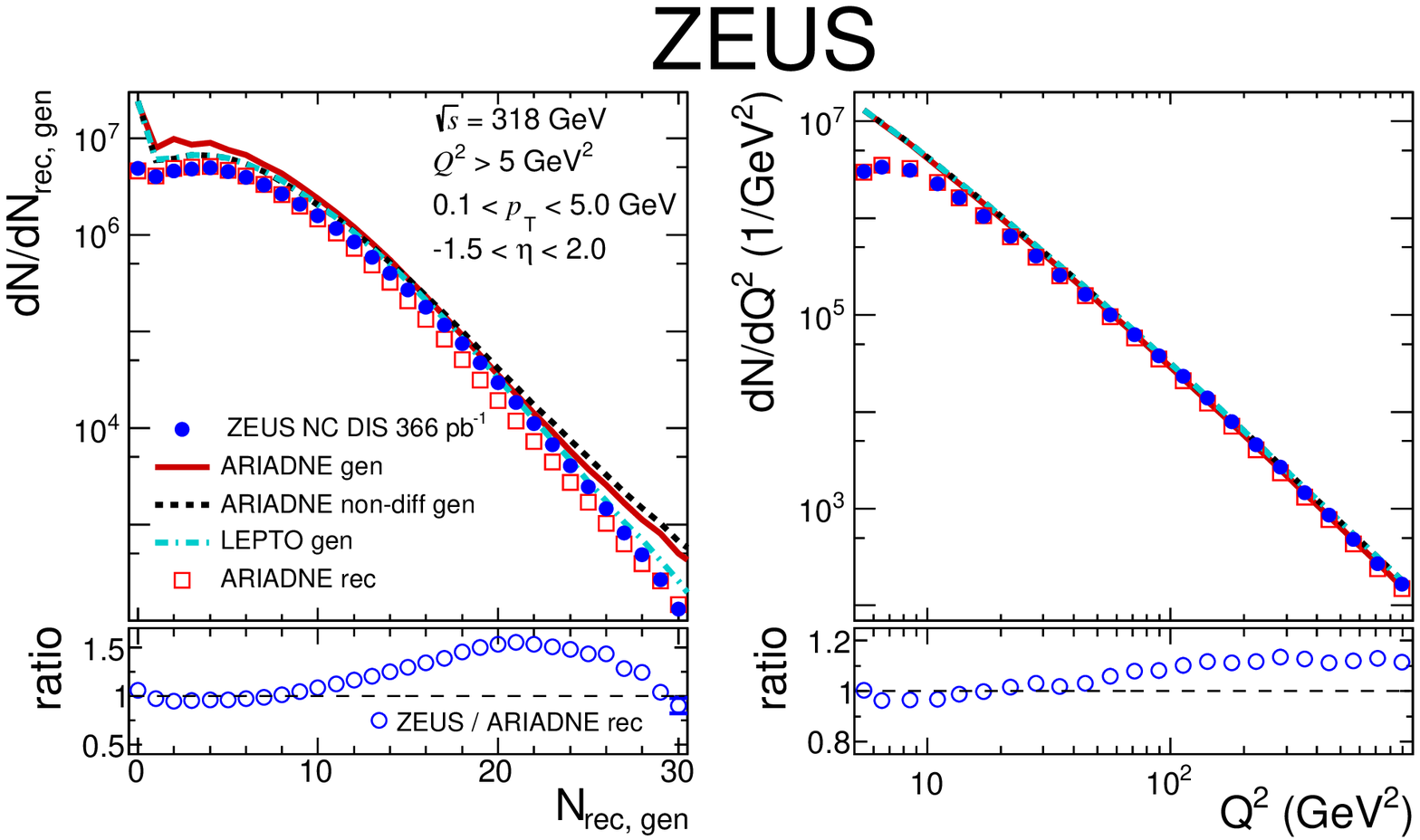}}
\put(60,170){(a)}
\put(290,170){(b)}
\end{picture}
  \caption{
  Reconstructed distributions of (a) multiplicity and (b) $Q^2$ in data compared to LEPTO and ARIADNE model predictions.
  The reconstructed Monte Carlo distributions are normalised to the total number of reconstructed events in data.
  Generator-level distributions are also shown using the same scale factors as for the reconstructed distributions.
  The normalisation procedure for LEPTO follows that for ARIADNE.
  The statistical uncertainties are smaller than the marker size.}
  \label{fig:MultQ2}
\end{figure}

To validate the extraction of efficiency corrections from ARIADNE and LEPTO, the data are now compared to model predictions at reconstruction level.  
The distributions of \Nrec and $Q^2$ are shown in Fig.~\ref{fig:MultQ2}.
In addition, Fig.~\ref{fig:MultQ2} shows distributions at generator level.
For this, Monte Carlo events were selected based on $Q^2$, $E_e$, $\theta_e$, and $E-p_Z$, which were calculated using initial- and final-state electron momenta.
These quantities were required to be in the same intervals as used at reconstruction level.
Generated primary particles were selected from the same kinematic \pT and $\eta$ intervals as at reconstruction level without a matching constraint.

Figure~\ref{fig:MultQ2}(a) shows the distributions of track multiplicity, where $N_{\textrm{gen}}$ denotes the number of selected generated tracks for either ARIADNE or LEPTO, and \Nrec denotes the number of selected reconstructed tracks in either data or Monte Carlo.
Figure \ref{fig:MultQ2}(b) shows the equivalent distributions in $Q^2$.
The reconstructed \Nrec and $Q^2$ distributions as predicted by ARIADNE are compatible with the data to within about 10\%, except for the high-\Nrec region, where the discrepancy is about 50\%.
The mean value of \Nrec is about 5 and the mean value of $Q^2$ is about $30\GeV^2$.

\begin{figure}[ht]
  \centering
\begin{picture}(450,400)
\put(0,0){\includegraphics[width=\textwidth]{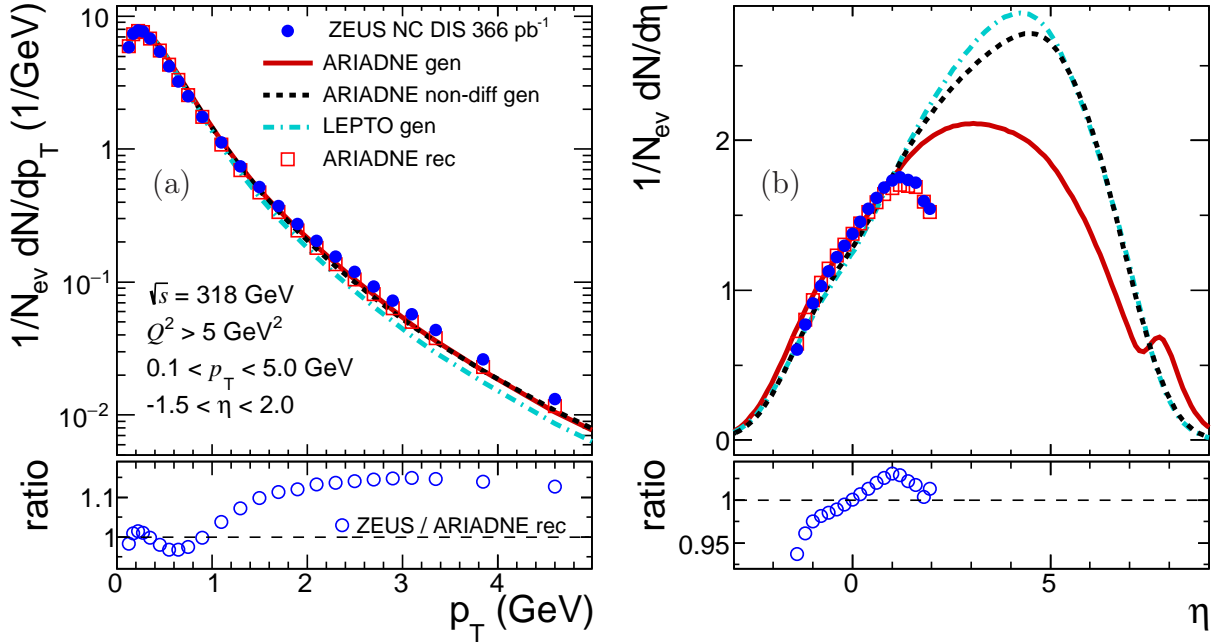}}
\put(60,170){(a)}
\put(290,170){(b)}
\end{picture}
  \caption{Reconstructed distributions of (a) \pT and (b) $\eta$ in data compared to LEPTO and ARIADNE.
  The reconstructed distributions are first normalised by their respective total number of events, $N_{\textrm{ev}}$.
  The generator-level predictions of ARIADNE are normalised to reconstructed ARIADNE at (a) $\pT=0.1\GeV$~and at (b) $\eta=0$.
  The other model predictions have been normalised by the same factor (1.3).
  The kink in the ARIADNE prediction near $\eta=8$ arises from the contribution of diffractive events where the incoming proton remains intact.
  The other details are as in Fig.~\ref{fig:MultQ2}.}
  \label{fig:PtEta}
\end{figure}

Reconstructed \pT and $\eta$ track distributions are shown in Fig.~\ref{fig:PtEta}.
The reconstructed single-particle distributions in ARIADNE are compatible with the data to within about 10\% except for the high-\pT region, where the discrepancy is about 15\%.
Owing to the asymmetric electron and proton beam energies and the occurrence of a beam remnant in the proton direction, particle production is expected to peak in the forward direction near $\eta=4$.

\begin{figure}[ht]
  \centering
  \begin{picture}(400,400)
  \put(0,0){\includegraphics[width=0.75\textwidth]{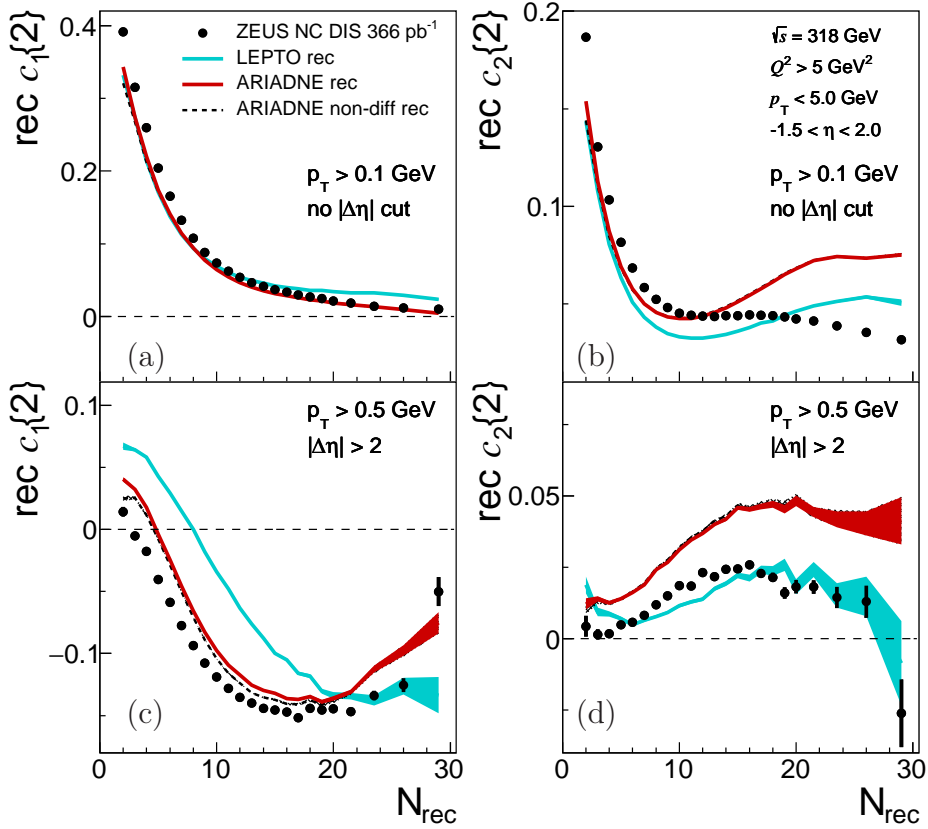}}
  \put(45,175){(a)}
  \put(215,175){(b)}
  \put(45,40){(c)}
  \put(215,40){(d)}
  \end{picture}
  \caption{Reconstructed \cn{1} and \cn{2} versus \Nrec.  The (a) and (b) panels represent the kinematic intervals given by $\pT>0.1\GeV$~and no \deta cut.  The (c) and (d) panels are further constrained by $\pT>0.5\GeV$~and $\deta>2$.  
  The predictions from ARIADNE, ARIADNE non-diffractive, and LEPTO are shown.
  The ARIADNE non-diffractive prediction is often hidden by the band of the full ARIADNE prediction.
  Statistical uncertainties are shown with vertical lines for ZEUS data and with bands for the Monte Carlo predictions.}
  \label{fig:rec_c12_Nrec}
\end{figure}

The reconstructed two-particle correlations \cn{1} and \cn{2} versus \Nrec in data and Monte Carlo are shown in Fig.~\ref{fig:rec_c12_Nrec}.
Figures \ref{fig:rec_c12_Nrec}(a) and (b) represent the kinematic intervals given by $\pT>0.1\GeV$~and no \deta cut, while (c) and (d) represent $\pT>0.5\GeV$~and $\deta>2$.
From the comparison of the full ARIADNE distributions with the distributions predicted by the non-diffractive component only, it is clear that the impact of diffraction on these distributions is small.
Thus, ARIADNE and LEPTO can be compared to the data on an equal footing.
It is clear that ARIADNE describes \cn{1} better than LEPTO in panels (a) and (c) while the opposite is true with \cn{2} in panel (d) with a \deta and stronger \pT cut.
For \cn{2}, in panel (b), which corresponds to the full kinematic interval, neither model fully satisfactorily describes the data.

\begin{figure}[ht]
  \centering
  \begin{picture}(400,400)
  \put(0,0){\includegraphics[width=0.75\textwidth]{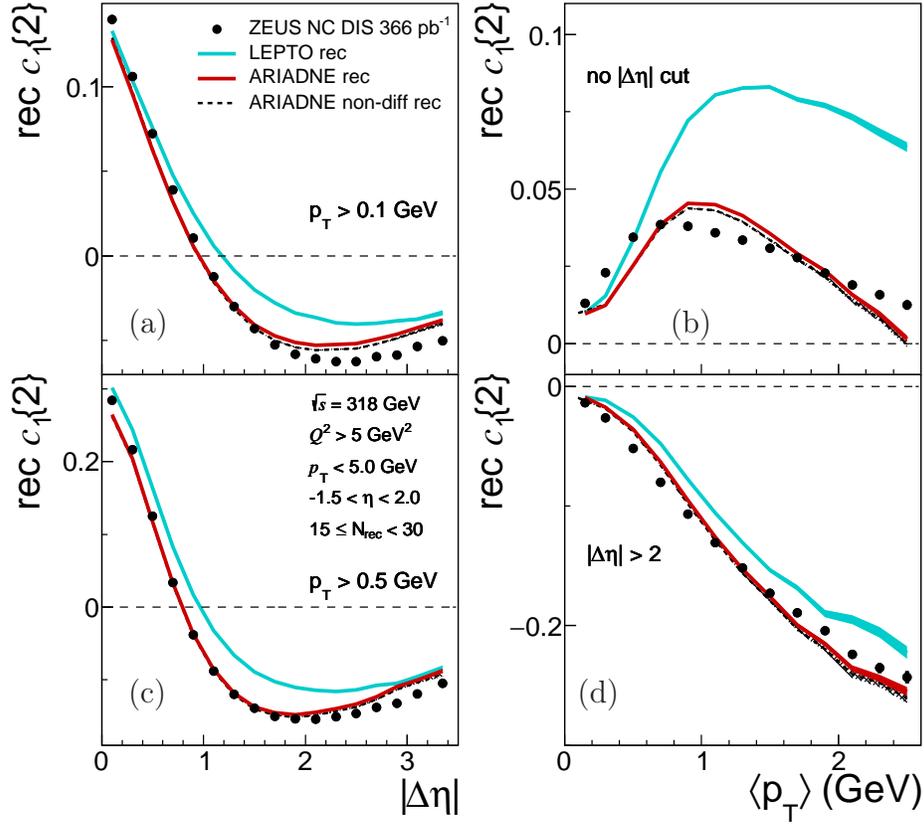}}
  \put(45,185){(a)}
  \put(250,185){(b)}
  \put(45,45){(c)}
  \put(215,45){(d)}
  \end{picture}
  \caption{Reconstructed \cn{1} versus \deta and \meanpt.  The (a) and (b) panels represent the kinematic intervals given by $\pT>0.1\GeV$~and no \deta cut.  The (c) and (d) panels are further constrained to $\pT>0.5\GeV$~and $\deta>2$, respectively. 
  The other details are as in Fig.~\ref{fig:rec_c12_Nrec}.}
  \label{fig:rec_c1_deta_mpt}
\end{figure}

\begin{figure}[ht]
  \centering
  \begin{picture}(400,400)
  \put(0,0){\includegraphics[width=0.75\textwidth]{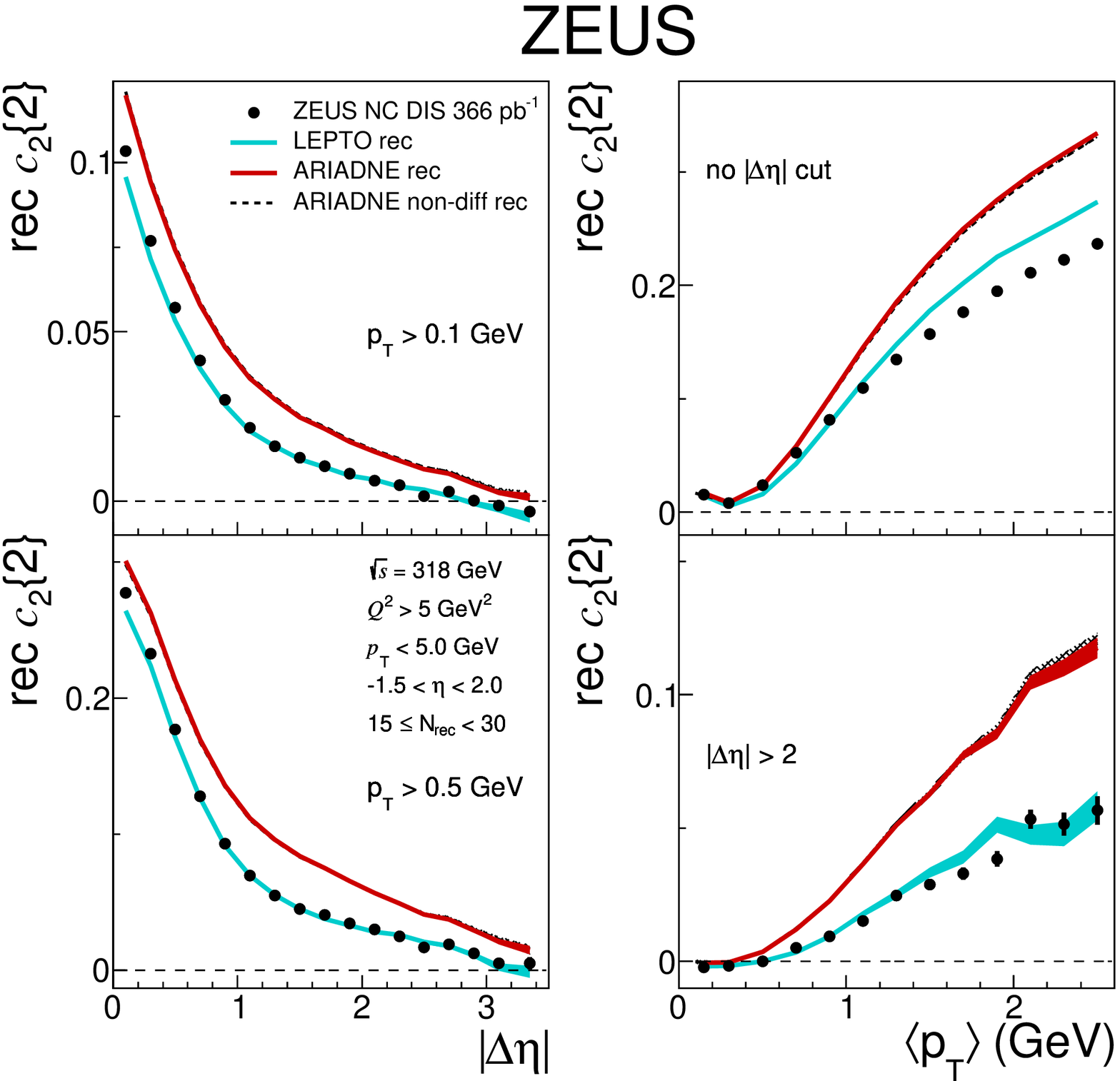}}
  \put(45,190){(a)}
  \put(215,190){(b)}
  \put(45,55){(c)}
  \put(215,55){(d)}
  \end{picture}
  \caption{Reconstructed \cn{2} as a function of \deta and \meanpt.  The other details are as in Fig.~\ref{fig:rec_c1_deta_mpt}.}
  \label{fig:rec_c2_deta_mpt}
\end{figure}

Reconstructed two-particle correlations \cn{1} as a function of \deta and \meanpt are shown in Fig.~\ref{fig:rec_c1_deta_mpt}.
It is clear that ARIADNE describes the data much better than LEPTO in all of the kinematic intervals shown.
In contrast, \cn{2} as a function of \deta and \meanpt is described much better by LEPTO as shown in Fig.~\ref{fig:rec_c2_deta_mpt}.
In all cases the data are described by at least one of the two Monte Carlo models reasonably enough, such that the efficiency corrections can be derived reliably.

\section{Efficiency corrections}
\label{Sec:Efficiency}

The measurement of two-particle correlations can be affected by non-uniform particle-tracking efficiency.
The single- and two-particle efficiencies were estimated by comparing the number of primary particles or pairs as generated with ARIADNE to the corresponding reconstructed numbers.
The single-particle efficiencies were extracted differentially in \pT, $\eta$, $\varphi$, charge, and data-taking period.
Two-particle efficiencies, which characterise the degree to which two tracks close in $\varphi$ can be distinguished in the CTD, were extracted differentially in $\Delta\varphi$, $\deta$, \Nch, and relative charge.

Corrections for non-uniform efficiency were applied using two types of weights.  They were extracted in two steps from Monte Carlo event samples. 
In the first step, the single-particle tracking efficiencies were calculated as the ratio of the number of reconstructed to generated particles passing the track-selection criteria.
The weight for particle $i$, $w_i$, is the inverse of the single-particle tracking efficiency.
Such weights are valid provided that there are no regions void of reconstructed particles (holes), which is the case for the chosen kinematic interval. 
Projected against \pT, the typical variation of $w_i$ from its mean value is about $5\%$ at high \pT and $15\%$ at low \pT, where secondary contamination becomes larger.
The typical variation of $w_i$ projected against $\varphi$ is about $5\%$.
The true number of charged primary particles within the fiducial region in a given event was estimated with a weighted sum over the reconstructed tracks passing the track-selection criteria, \Nrec:
\begin{equation}
    \Nch = \sum_i^{\Nrec} w_i.
    \label{eq:Nch}
\end{equation}

In a second pass over the Monte Carlo events, the $w_i$ weights were applied to the reconstructed particles and two-particle reconstruction efficiencies were calculated as a function of $\Delta\varphi$.
The ratio of the number of generated to reconstructed pairs passing the track-selection criteria forms the second weight \wdphi.
The typical variation of \wdphi is about $10\%$ and is largest for same-sign pairs with $|\Delta\varphi| < 0.3$ radians.

\section{Analysis method}
\label{Sec:Analysis}

The two-particle correlation functions measured in this analysis are defined by
\begin{equation}
  c_{n}\{2\} = \left. \sum_{e}^{N_{\textrm{ev}}} \left[ \sum_{i, j > i}^{\Nrec} w_{ij}
  \cos{ [ n (\varphi_{i} -\varphi_{j}) ] } \right]_e \; \right/ \sum_e^{N_{\textrm{ev}}} \left[ \sum_{i, j > i}^{\Nrec} w_{ij} \right]_e,
  \label{eq:cn}
\end{equation}
where $\varphi_i$ and $\varphi_j$ are the azimuthal angles of the two particles.
The first sum over $e$ is performed for all events, $N_{\textrm{ev}}$, and the sums over $i$ and $j$ run over all selected charged particles in the event with multiplicity \Nrec.
The pair-correction factor for non-uniform acceptance is given by $w_{ij} = w_i w_j \wdphi$, which is normalized (See Eq.~\ref{eq:cn}) in the determination of \cn{n}.

Two-particle correlations are also reported in a two-dimensional form, which is defined as:
\begin{equation}
    C(\Delta\eta,\Delta\varphi) = \frac{ S(\Delta\eta,\Delta\varphi) }{ B(\Delta\eta,\Delta\varphi) },
\end{equation}
where \mbox{$S(\Delta\eta,\Delta\varphi) = N^{\textrm{same}}_{\textrm{pairs}}(\Delta\eta,  \Delta\varphi)$} and \mbox{$B(\Delta\eta,\Delta\varphi) = N^{\textrm{mixed}}_{\textrm{pairs}} (\Delta\eta,  \Delta\varphi)$} are the number of pairs for the signal and background distributions, respectively.
These pair distributions were formed by taking the first particle from a given event and the other from either the same event or a different event (mixed) with similar values of \Nrec and vertex $Z$ position. 
The $S$ distribution was corrected with $w_{ij}$, while $B$ was corrected with $w_i w_j$.
Both distributions were symmetrised along $\Delta\eta$ and then individually normalised to unity before division.

\section{Systematic uncertainties}
\label{Sec:Systematics}

In principle, the application of the efficiency corrections as defined in Eq.~\ref{eq:cn} to the reconstructed Monte Carlo data should recover the distributions of \cn{n} at generator level. 
However, residual differences persist.
This is called Monte Carlo non-closure.
Qualitatively, the Monte Carlo non-closure was observed to be similar for ARIADNE and LEPTO.
Quantitatively, differences were observed, because the models predict different event configurations to which the detector responds differently.
For the results, the Monte Carlo non-closure values from ARIADNE, $\dncAri$, and LEPTO, $\dncLepto$, were averaged, $\dnc$, and are quoted as a signed separate uncertainty.
The typical values of this uncertainty on \cn{n} versus \Nch without a cut
on \deta are $<15\%$.

Further systematic uncertainties were estimated by comparing the correlations obtained with the default event- and track-selection criteria to those obtained with varied settings.
The difference between the \cn{n} results obtained with the default and the varied settings was assigned as the systematic uncertainty.
The sources of systematic uncertainty that were considered are given below (with the typical values of the uncertainty on \cn{n} versus \Nch without a cut on \deta):
\begin{itemize}
%\item Monte Carlo non-closure ($<15\%$);
\item secondary-particle contamination:  The default analysis used $\textrm{DCA}_{XY,Z}<2$ cm, while for the variation $\textrm{DCA}_{XY,Z}< 1$ cm was used.  The uncertainty was symmetrised ($<10\%$);
\item efficiency-correction uncertainty due to the choice of Monte Carlo generator: The default analysis used ARIADNE, while for the variation, LEPTO was used.  The uncertainty was largest at high \Nch ($<10\%$);
\item consistency of \cn{n} from events with different primary-vertex positions, $V_Z$: The default analysis used  $|V_Z| < 30$~cm. For the variations either $-30 < V_Z < 0$ cm or $0 < V_Z < 30$ cm were selected.  The resulting deviations were weighted by their relative contribution ($<5\%$);
\item low-\pT tracking efficiency: The default simulation included the low-\pT track rejection, while for the variation it did not.  The uncertainty was assigned to be half of the difference between the default and varied procedure and was symmetrised ($<3\%$);
\item different data-taking conditions: The default analysis used all available data, while for the variations, separate data taking periods weighted by their relative contribution were used and the differences were added in quadrature and used as a symmetric uncertainty ($<2\%$);
\item DIS event-selection criteria: The chosen $E-p_Z$ interval, the scattered-electron polar angle, the neural-network identification probability, and the excluded entrance locations of the scattered electron in the CAL were found to have a negligible effect. 
\end{itemize}
Each variation was applied to ZEUS data as well as Monte Carlo data for the recalculation of efficiency corrections.
Positive and negative systematic uncertainties were separately summed in quadrature to obtain the total systematic uncertainty, $\dsyst$.
The values of each systematic uncertainty and the full information for the two models are also provided in Tables \ref{Tab:Table16}--\ref{Tab:Table41}.

\section{Results}
\label{Sec:Results}

Results are presented\footnote{The values are given in Tables
\ref{Tab:Table16}--\ref{Tab:Table41}.} in the kinematic region defined by:
\mbox{$Q^2 > 5\GeV^2$}, $E_e > 10\GeV$, $E-p_Z > 47\GeV$, $\theta_e > 1$
radian, and primary charged particles with \mbox{$-1.5 < \eta < 2.0$} and
\mbox{$0.1 < \pT < 5.0\GeV$}.
The kinematic intervals were chosen to avoid contributions from unwanted hard processes at very high \pT~and to provide good tracking efficiency.

\begin{figure}
  \centering
  \begin{picture}(500,400)
  \put(0,0){\includegraphics[width=\textwidth]{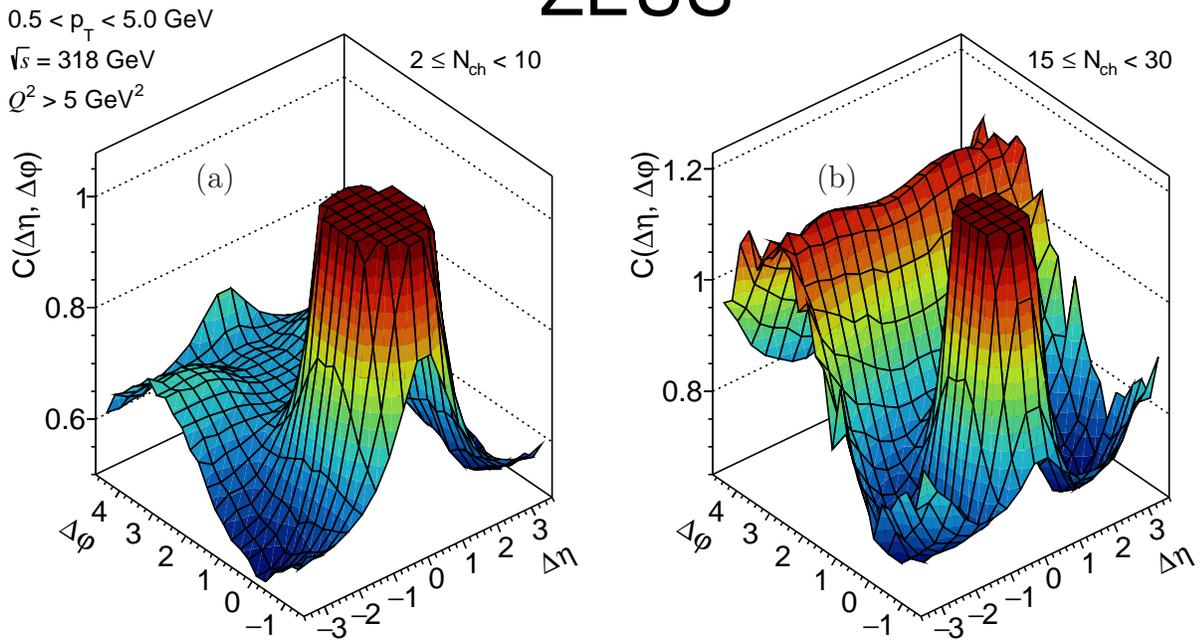}}
  \put(75,183){(a)}
  \put(310,183){(b)}
  \end{picture}
  \caption{Two-particle correlation \Cdetadphi for (a) low and (b) high \Nch.  The peaks near the origin have been truncated for better visibility of the finer structures of the correlation.  The plots were symmetrised along $\Delta\eta$.  No statistical or systematic uncertainties are shown.}
  \label{fig:C_deta_dphi}
\end{figure}

Figure~\ref{fig:C_deta_dphi} shows \Cdetadphi for low and high \Nch and for
particles with $0.5 < \pT < 5.0\GeV$.
For both ranges in \Nch, a dominant near-side ($\Delta\varphi \sim 0$) peak is seen at small $\Delta \eta$.
The displayed range in $C$ was truncated to illustrate better the finer structures of the correlation.
Also in both \Nch ranges, at $\Delta\varphi \sim \pi$ (away-side), a broad ridge-like structure is observed.
At low \Nch, a dip in this away-side ridge is visible, while at high \Nch it is more uniform.
There is no indication of a near-side ridge with or without the subtraction of \Cdetadphi at low \Nch~from that at high \Nch, which would be an indication of hydrodynamic collectivity. 
This is in contrast to what has been observed in high-multiplicity $pp$ and $p+Pb$ collisions \cite{Khachatryan:2010gv, Aad:2012gla, Abelev:2012ola}.
Similarly, an analysis of two-particle correlations in $e^+ e^-$ shows no indication of a near-side ridge~\cite{Badea:2019vey}. 

\begin{figure}[ht]
  \centering
  \begin{picture}(400,400)
  \put(0,0){\includegraphics[width=0.75\textwidth]{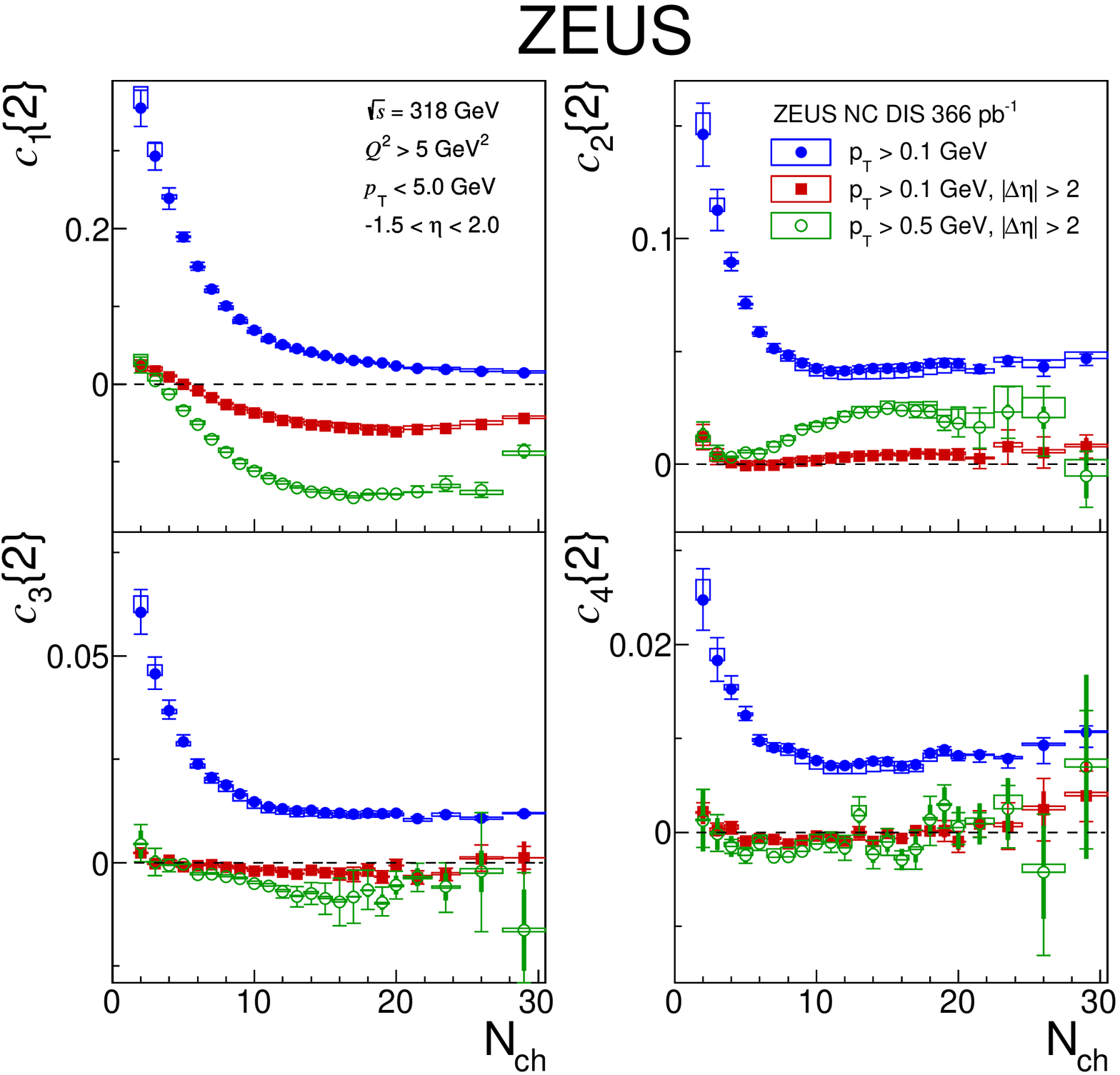}}
  \put(140,235){(a)}
  \put(310,235){(b)}
  \put(140,130){(c)}
  \put(310,130){(d)}
  \end{picture}
  \caption{Correlations \cn{n} as a function of \Nch~for (a) $n=1$, (b) $n=2$, (c) $n=3$, (d) $n=4$, with and without a rapidity separation, and for low- and high-\pT intervals.  The statistical uncertainties are shown as thick vertical lines although they are typically smaller than the marker size.  Systematic uncertainties from the Monte Carlo non-closure (see Section \ref{Sec:Systematics}) are shown as boxes.  
  The other systematic uncertainties are shown as a thin vertical capped line.}
  \label{fig:cn_Nch}
\end{figure}

Figure~\ref{fig:cn_Nch} shows the \Nch~dependence of the two-particle correlations \cn{n} for the first four harmonics, $n=1-4$.
Results are presented for the full ranges of \deta and \pT, and with a rapidity-separation condition, $\deta>2$, for $\pT > 0.1$ and $\pT > 0.5\GeV$.
Without a rapidity separation, the \cn{n} correlations are strongest and positive at low \Nch for all $n$, indicating that particles are preferentially emitted into the same hemisphere, as expected for the fragmentation of the struck parton.
This is largely absent for $\deta > 2$, indicating that \cn{n} at small multiplicities is dominated mostly by short-range correlations.
An alternative way to suppress short-range correlations is to use multiparticle correlations \cite{Borghini:2001vi} such as four-particle cumulants $c_n{\scriptstyle \{4\}}$, which explicitly removes them.
Owing to limited statistics, they were not studied here.

All \cn{n} correlations depend only weakly on \Nch for $\Nch>15$.
For $\deta>2$, \cn{1} and \cn{3} become negative, which is expected from the effects of global momentum conservation, e.g. back-to-back dijet-like processes with large eta separation between jets.
For $\pT>0.5\GeV$, \cn{1} (\cn{2}) exhibit stronger negative (positive) correlations than for $\pT>0.1\GeV$.

Similar conclusions can be drawn by comparing Fig.~\ref{fig:C_deta_dphi} to Fig.~\ref{fig:cn_Nch}.
For low \Nch and no \deta cut, the peak in Fig.~\ref{fig:C_deta_dphi} is the dominant structure from which $\cn{1} = \left< \left< \cos{\Delta\varphi} \right> \right> > 0$ and $\cn{2} = \left< \left< \cos{2\Delta\varphi} \right> \right> > 0$ are expected.
For large values of \Nch, \deta and \pT, the away-side ridge becomes the dominant structure, which leads to the pattern $\left< \left< \cos{2\Delta\varphi} \right> \right> > 0$ and $ \left< \left< \cos{\Delta\varphi} \right> \right> < 0$.
It can be seen that $|\cn{1}|$ is much larger than $|\cn{2}|$ at high \Nch and \deta.
This reflects that inclusive NC DIS events have a more directed than elliptic event topology.
This is in contrast to systems with larger interaction regions, where the positive magnitude of \cn{2} is much larger than the negative magnitude of \cn{1}\cite{Aamodt:2011by}, which is an expected signature of hydrodynamic collectivity. 

\begin{figure}[ht]
  \centering
  \begin{picture}(400,400)
  \put(0,0){\includegraphics[width=0.75\textwidth]{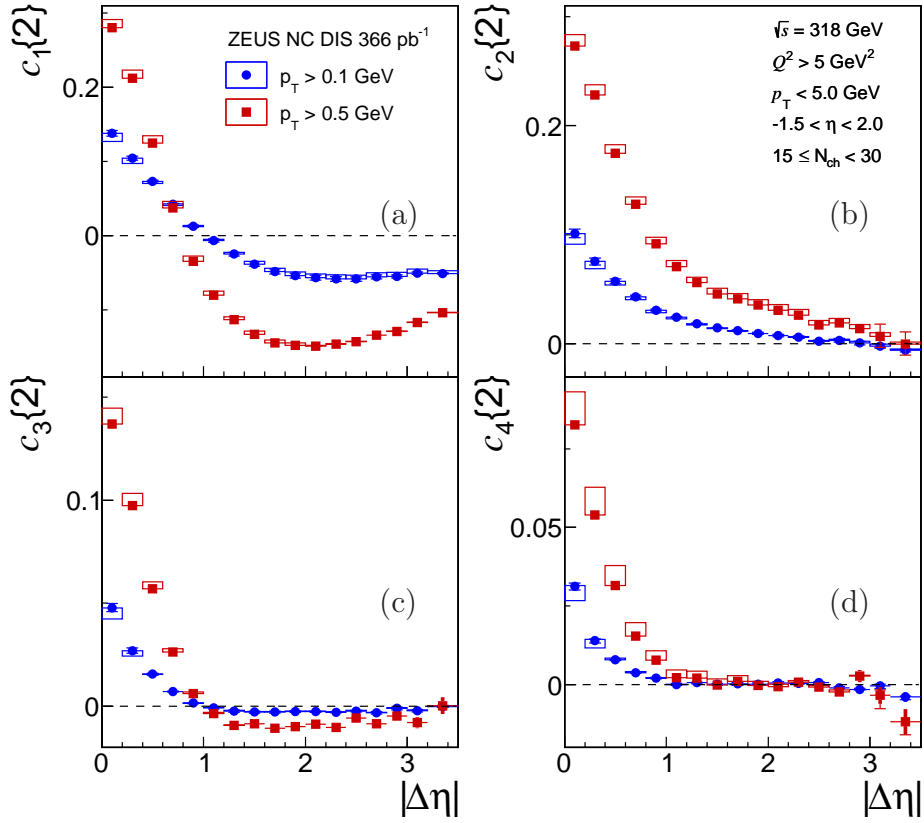}}
  \put(140,227){(a)}
  \put(310,227){(b)}
  \put(140,80){(c)}
  \put(310,80){(d)}
  \end{picture}
  \caption{Correlations \cn{n} for $15 \leq \Nch < 30$ as a function of \deta for (a) $n=1$, (b) $n=2$, (c) $n=3$, (d) $n=4$.  
  Two selections of transverse momentum intervals are shown: $\pT>0.1\GeV$~and $\pT>0.5\GeV$.
  The other details are as in Fig.~\ref{fig:cn_Nch}.}
  \label{fig:cn_dEta}
\end{figure}

Figure \ref{fig:cn_dEta} shows the two-particle correlations as a function of rapidity separation $\deta$.
Compared to results for $\pT>0.1\GeV$, the correlations with $\pT>0.5\GeV$~are more pronounced, as expected from particles in jet-like structures.
The mean values of \pT in the low- and high-\pT intervals are 0.6 and 1.0\GeV, respectively.
The correlations \cn{1} and \cn{3} have qualitatively similar dependence on $\deta$ but with different modulation strengths.
Both change sign near $\deta=1$, which shows that the short-range correlations extend up to about one unit of rapidity separation, after which the long-range effects, such as global momentum conservation, become dominant contributions to \cn{1} and \cn{3}.
Integrated for $\pT>0.1\GeV$, \cn{2} approaches zero for $\deta \gtrsim 2$.
Positive correlations observed in \cn{2} for $\pT>0.5\GeV$~extend out to $\deta\sim3$.
%The correlations from all harmonics except for the first are consistent with zero for $\deta > 3$.

\begin{figure}[ht]
  \centering
  \begin{picture}(400,200)
  \put(0,0){\includegraphics[width=0.75\textwidth]{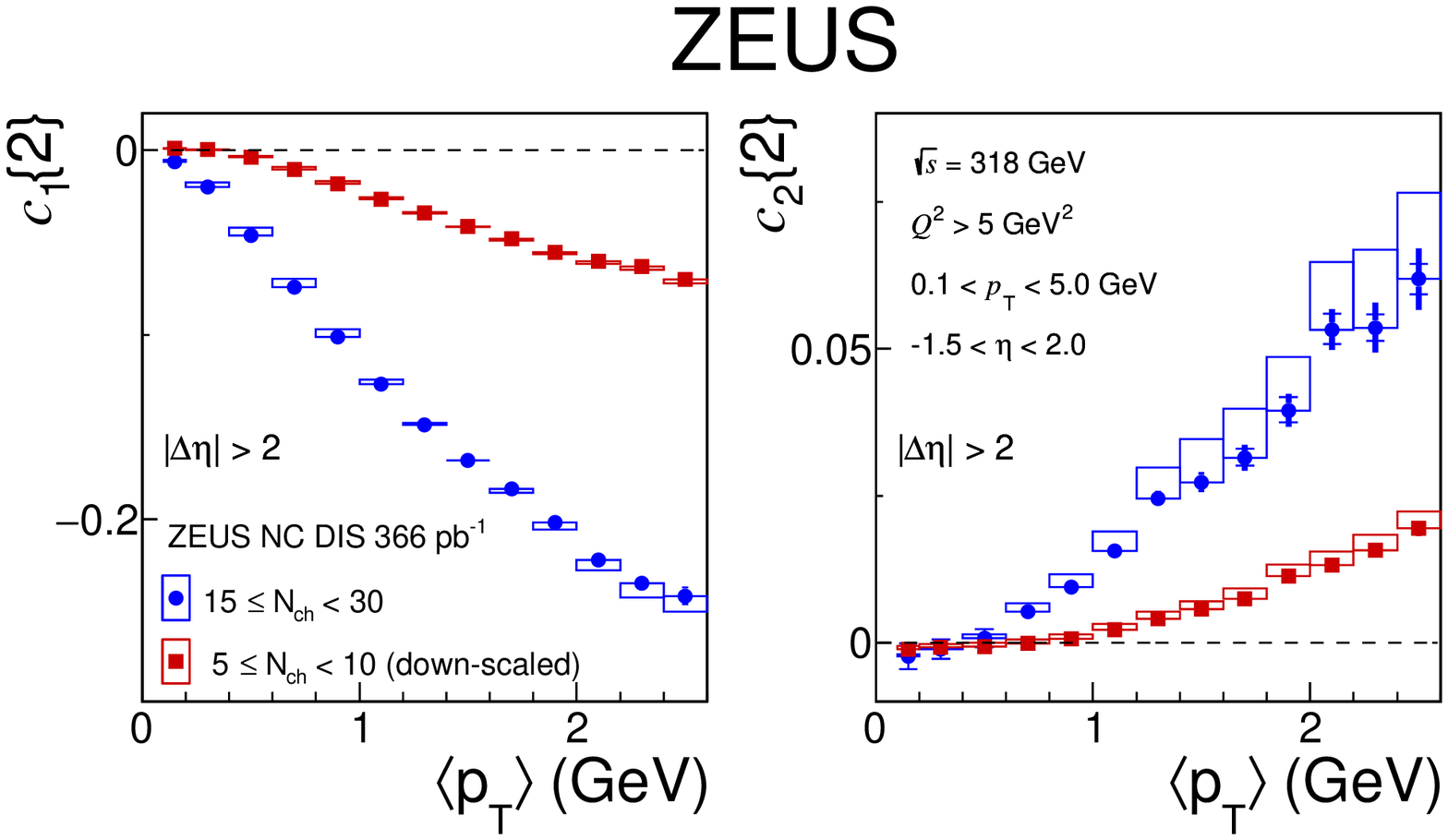}}
  \put(130,152){(a)}
  \put(300,152){(b)}
  \end{picture}
  \caption{Correlations \cn{1} and \cn{2} as a function of $\meanpt$ for (a) low and (b) high $\Nch$.  Low $\Nch$ correlations are down-scaled by a factor of $\left< \Nch \right>_{low} / \left< \Nch \right>_{high}$ as explained in the text.
  The other details are as in Fig.~\ref{fig:cn_Nch}.}
  \label{fig:c12_mpt}
  \end{figure}

In Fig.~\ref{fig:c12_mpt}, \cn{1} and \cn{2} are plotted versus $\meanpt$ with $\deta > 2$ in low- and high-multiplicity regions. 
The third and fourth harmonic correlation functions have much larger statistical uncertainties and are therefore not shown. 
Correlations at low $\Nch$ were down-scaled by the factor $\left< \Nch \right>_{\textrm{low}} / \left< \Nch \right>_{\textrm{high}} = 0.4$, where $\left< \Nch \right>_{\textrm{low}}$ ($\left< \Nch \right>_{\textrm{high}}$) = 6.7 (16.8).
Studies in heavy-ion collisions suggest that correlations unrelated to hydrodynamic collectivity contribute to \cn{n} as 1/$\Nch$~\cite{Adams:2004wz,Abelev:2012di}.
Applying the scaling factor provides a better means to compare and investigate the possible collective effects, which enter each multiplicity interval differently, and investigate if there is an excess of the correlations at high multiplicities.
For both \cn{1} and \cn{2}, the correlation strength grows with increasing $\meanpt$ up to a few \GeV, which is universally observed in all collision systems \cite{Abelev:2013vea, Khachatryan:2010gv, Abelev:2012ola, Aad:2012gla, Aad:2015gqa, Adare:2013piz, Adare:2015ctn}.
Despite the observed excess of correlation strength for \cn{2} at high compared to low multiplicity, an even stronger excess is observed for \cn{1}, which, as described above, is dominated by dijet-like processes.
This suggests that the 1/$\Nch$ scaling inspired by observations in heavy-ion collisions may not be appropriate for \ep scattering.

Comparisons of \cn{2} at low and high multiplicity, as well as fits to \Cdetadphi, have been performed at RHIC and LHC.
The laboratory rapidity window used in the analysis presented here is located about 2--5 units away from the peak of the proton fragmentation region at $\eta \approx 4$ (Fig.~\ref{fig:PtEta}). The LHC measurements in \pp collisions were made in between two wide fragmentation peaks which are separated by about 4 units~\cite{ALICE:2012xs,Khachatryan:2015jna,Adam:2015pza}.
Despite this difference in rapidity coverage, the typical magnitudes of \cn{2} are compared.
The value of \cn{2} at RHIC \cite{Adare:2013piz, Adare:2015ctn} and LHC \cite{Khachatryan:2010gv, Aad:2012gla, Abelev:2012ola, Abelev:2013vea, Aad:2015gqa} lies in the range $0.002$--$0.01$ at $\pT\approx1\GeV$~\cite{Aad:2015gqa}.
At a similar \pT~value, the corresponding difference between the central values of \cn{2} at low and high multiplicity in Fig.~\ref{fig:c12_mpt} is about 0.01. 
The further understanding of the similarity of the \cn{2} excess observed in both \ep and \pp, together with the much larger \cn{1} excess relative to that for \cn{2} in \ep, would require a consistent modelling of multi-particle production in both collision systems.

\begin{figure}[ht]
  \centering
  \begin{picture}(400,400)
  \put(0,0){\includegraphics[width=0.75\textwidth]{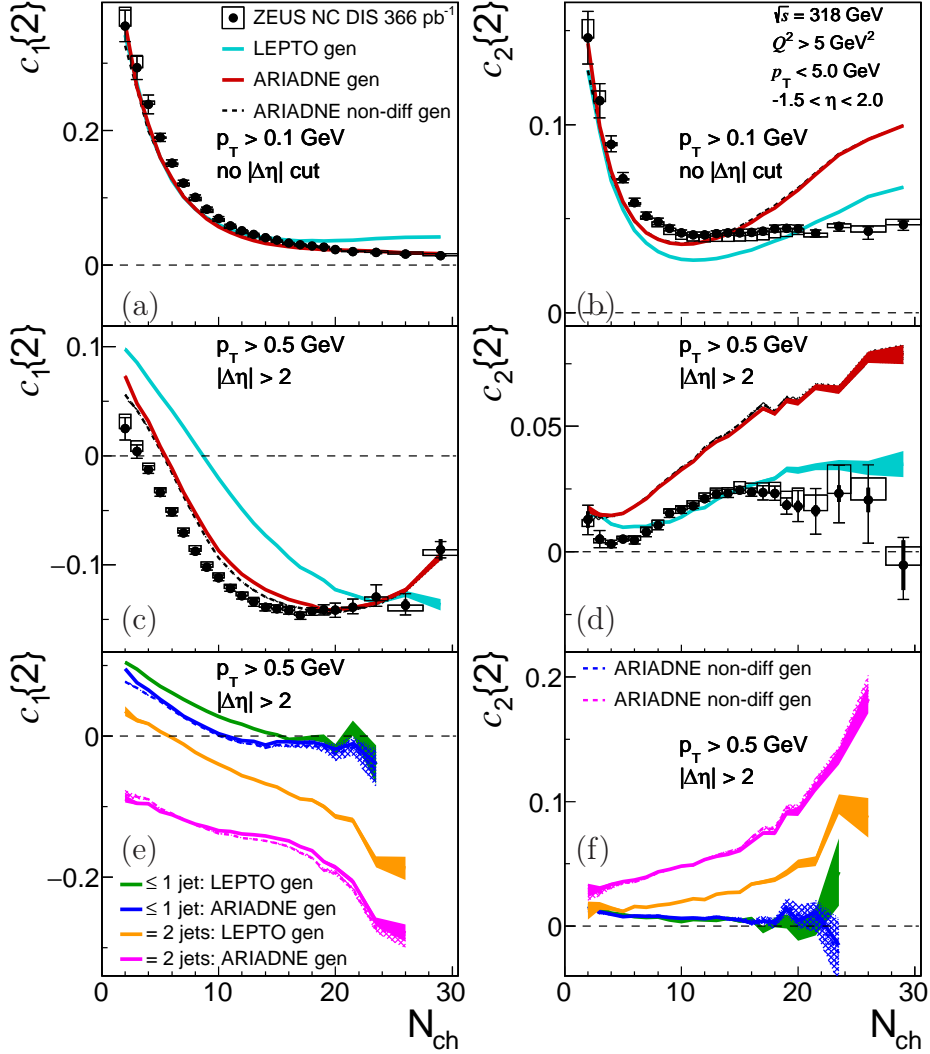}}
  \put(42,279){(a)}
  \put(214,279){(b)}
  \put(42,160){(c)}
  \put(214,160){(d)}
  \put(42,73){(e)}
  \put(214,73){(f)}
  \end{picture}
  \caption{Correlations \cn{1} and \cn{2} with and without a rapidity separation as a function of $\Nch$ compared to the predictions from Monte Carlo event generators.  Correlations measured in the full kinematic interval are shown in (a) and (b), while (c) and (d) represent the interval given by $\pT>0.5\GeV$ and $\deta>2$.  Panels e) and f) separate out contributions from events with $\leq 1$ jet and 2 jets.  The correlation from a non-diffractive component in ARIADNE is shown with dashed lines.
  The other details are as in Fig.~\ref{fig:cn_Nch}.}
  \label{fig:cn_MC}
  \end{figure}

\begin{figure}[ht]
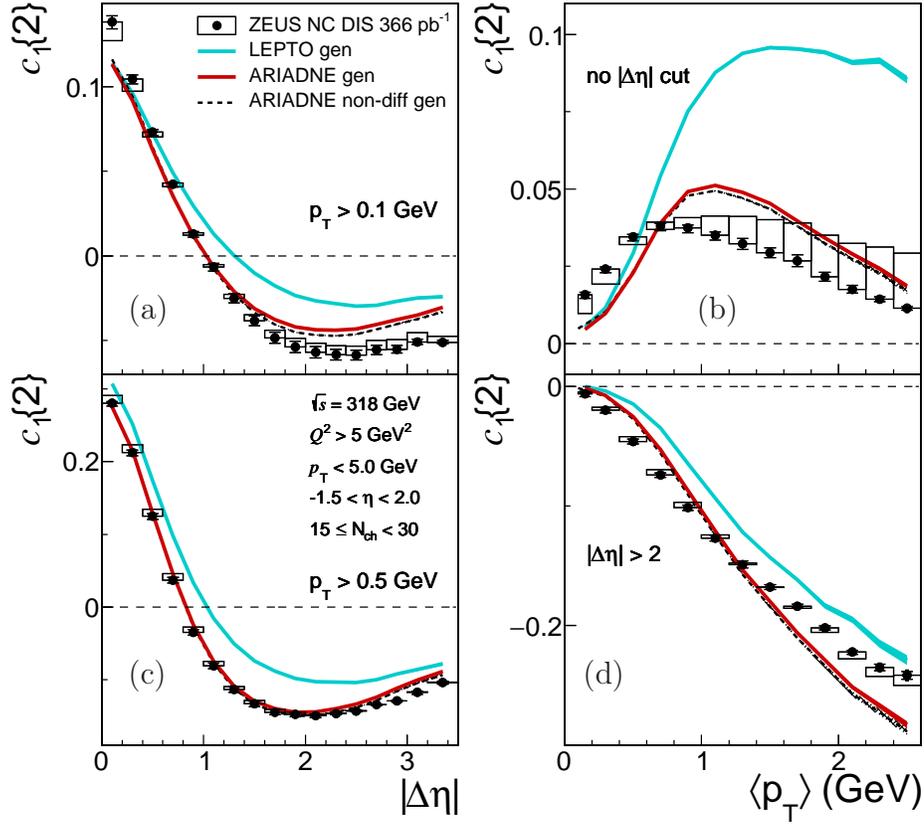

  \centering

  \caption{Correlations \cn{1} as a function of \deta~and $\meanpt$ for $15 \leq \Nch < 30$. Also shown are generated correlations from Monte Carlo models.  Correlations measured in the full kinematic interval are shown in (a) and (b), while (c) and (d) are further constrained by $\pT>0.5\GeV$ and $\deta>2$, respectively.
  The other details are as in Fig.~\ref{fig:cn_Nch}.}
  \label{fig:c1_deta_mpt}
  \end{figure}

\begin{figure}[ht]
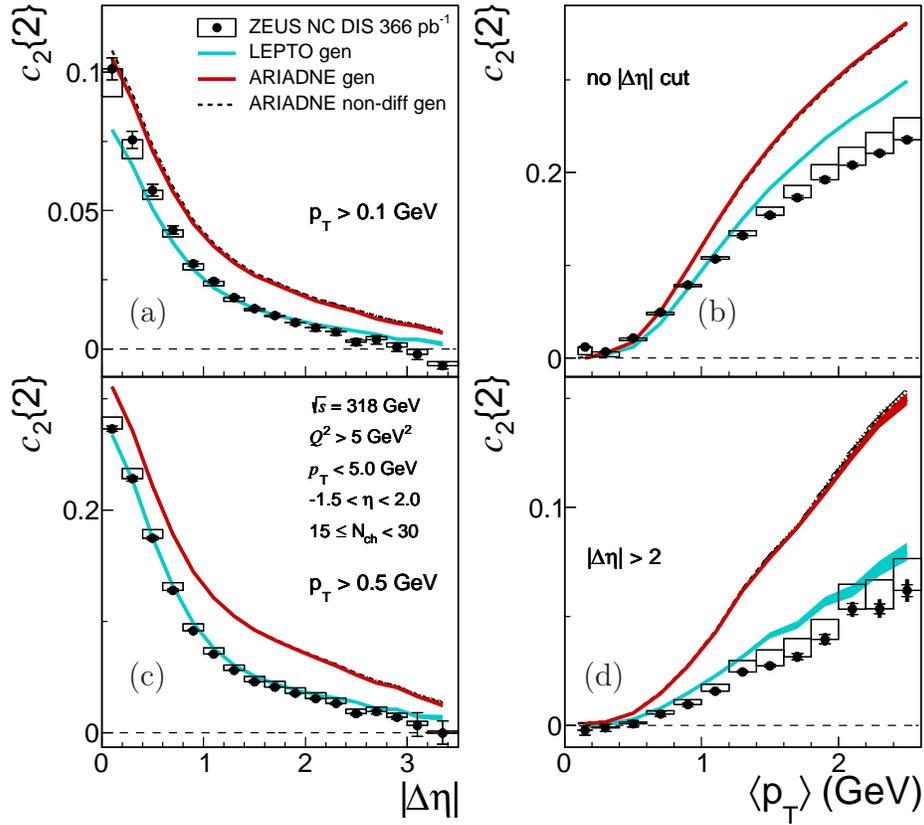

  \centering
  %
  \caption{Correlations \cn{2} as a function of \deta~and $\meanpt$ for $15 \leq \Nch < 30$. Also shown are the predictions from Monte Carlo event generators.
  The other details are as in Fig.~\ref{fig:c1_deta_mpt}.}
  \label{fig:c2_deta_mpt}
  \end{figure}

The generated correlations in LEPTO and ARIADNE are compared to the measured correlations \cn{1} and \cn{2} in Figs.~\ref{fig:cn_MC}--\ref{fig:c2_deta_mpt}.
In Fig.~\ref{fig:cn_MC}, generated correlations are compared to \cn{1} and \cn{2} versus $\Nch$(a) and (b) without and (c) and (d) with a \deta cut. 
The ARIADNE prediction is shown with and without a diffractive component.
Figures \ref{fig:cn_MC} (e) and (f) show the expectations from the models for $\leq 1$ and 2 jets.  
Massless jets were reconstructed from generated hadrons using the $k_T$ algorithm  \cite{Catani:1993hr} with $\Delta R=1$,
rapidity less than 3, and at least 2 GeV of transverse energy in the laboratory frame.
Figures \ref{fig:cn_MC} (e) and (f) confirm that dijet-like processes are responsible for the large values of $|\cn{1}|$ and $|\cn{2}|$.
The models are able to reproduce the qualitative features of the data but do not give a quantitative description in certain regions.
Both LEPTO and ARIADNE predict an increase of integrated \cn{2} at high \Nch, which the data do not show.

The correlations projected against \deta and \meanpt in Figs.~\ref{fig:c1_deta_mpt} and \ref{fig:c2_deta_mpt} confirm the observation at reconstruction level (Section \ref{Sec:RawData}) that ARIADNE describes \cn{1} better than LEPTO while the opposite is true for \cn{2}.
In particular, it is clear that long-range correlations ($\deta \gtrsim 2$) are underestimated by LEPTO for the first harmonic while they are overestimated by ARIADNE for the second harmonic.
The growth of \cn{2} correlations with $\meanpt$ is greatly overestimated by the ARIADNE model.

\section{Summary and outlook}
\label{Sec:Summary}

Two-particle azimuthal correlations have been measured with the ZEUS detector at HERA in neutral current deep inelastic \ep scattering at $\sqrt{s}=318\GeV$, using an integrated luminosity of $366\pm7\,\mathrm{pb}^{-1}$.
The kinematic region of the selected primary charged particles in the
laboratory frame is $0.1 < \pT < 5.0\GeV$~and $-1.5 < \eta < 2.0$.
The DIS scattered electron was constrained to have a polar angle greater than 1 radian relative to the proton beam direction, with energy larger than 10\GeV, $Q^2 > 5\GeV^2$.
The events were required to have $E-p_Z > 47\GeV$.

The correlations were measured for event multiplicities up to six times larger than the average $\left<\Nch\right>\approx5$.
There is no indication of a near-side ridge in \Cdetadphi.
Strong long-range anti-correlations are observed with \cn{1} as expected from global momentum conservation.
For $\pT > 0.5\GeV$, the observed anti-correlations in \cn{1} are stronger than the correlations in \cn{2}, which indicates that they originate from hard processes and not the collective effects that characterise RHIC and LHC data at high multiplicities.

Models of DIS, which are able to reproduce distributions of $Q^2$ and single-particle spectra, are able to qualitatively describe two-particle correlations but do not describe all distributions quantitatively.
In particular, LEPTO provides a better description of \cn{2}, while ARIADNE describes \cn{1} better.

The measurements demonstrate that the collective effects recently observed at RHIC and LHC are not observed in inclusive NC DIS collisions. 
Future studies with photoproduction are expected to shed light on the evolution of the multi-particle production mechanism from DIS to hadronic collisions, where the size of the interaction region changes from a fraction of a femtometer to femtometers.

\clearpage

% ----------------------------------------------------------------------------
%       Mandatory acknowledgements. You may add your buddies to it.
% ----------------------------------------------------------------------------
\section*{Acknowledgements}
\label{sec-ack}

\Zacknowledge

\vfill\eject

%------------------------------------------------------------------------------
%       Bibliography
%------------------------------------------------------------------------------

{
\ifzeusbst
  \ifzmcite
     \bibliographystyle{bst/l4z_default3}
  \else
     \bibliographystyle{bst/l4z_default3_nomcite}
  \fi
\fi
\ifzdrftbst
  \ifzmcite
    \bibliographystyle{bst/l4z_draft3}
  \else
    \bibliographystyle{bst/l4z_draft3_nomcite}
  \fi
\fi
\ifzbstepj
  \ifzmcite
    \bibliographystyle{bst/l4z_epj3}
  \else
    \bibliographystyle{bst/l4z_epj3_nomcite}
  \fi
\fi
\ifzbstjhep
  \ifzmcite
    \bibliographystyle{bst/l4z_jhep3}
  \else
    \bibliographystyle{bst/l4z_jhep3_nomcite}
  \fi
\fi
\ifzbstnp
  \ifzmcite
    \bibliographystyle{bst/l4z_np3}
  \else
    \bibliographystyle{bst/l4z_np3_nomcite}
  \fi
\fi
\ifzbstpl
  \ifzmcite
    \bibliographystyle{bst/l4z_pl3}
  \else
    \bibliographystyle{bst/l4z_pl3_nomcite}
  \fi
\fi
{\raggedright
\bibliography{bib/our_references.bib,%
              bib/l4z_zeus.bib,%
              bib/l4z_conferences.bib,%
              bib/l4z_articles.bib,%
              bib/l4z_temporary.bib,%
              bib/l4z_misc.bib}} 
}
\vfill\eject

%\bibliography{references}{}
%\bibliographystyle{unsrt}

%------------------------------------------------------------------------------
%       Tables
%------------------------------------------------------------------------------
%\newcommand{\TO}{\hspace*{-1.5ex}--\hspace*{-1.5ex}}
%\newcommand{\BR}{\hspace*{1.2ex}}
%\newcommand{\B}{$\hspace*{1.2ex}$}
%\newcommand{\sta}{\mathrm{(stat.)}}
%\newcommand{\sys}{\mathrm{(sys.)}}
%\renewcommand{\strut}{\rule[-1.4ex]{0ex}{4ex}}
%\newcommand{\strutb}{\rule[-0.4ex]{0ex}{3ex}}
%-------------------------------------------------------------------------------
%       Results
%------------------------------------------------------------------------------
\newpage

% Data tables
%\input{dataTables.tex}

\clearpage

% Figures


% Data tables
%cn vs Nch

\clearpage
\begin{small}
\begin{center}
\renewcommand*{\arraystretch}{1.2}
\begin{longtable}{| c | c | c | c | c | c | c |}
\hline

\hline\multicolumn{1}{|c|}{$\Nch$} & \multicolumn{1}{|c|}{$c_1\{2\}$} & \multicolumn{1}{|c|}{$\delta_{\textrm{stat}}$} & \multicolumn{1}{|c|}{$\delta_{\textrm{syst}}$} & \multicolumn{1}{|c|}{$\delta_{\textrm{nc}}$} & \multicolumn{1}{|c|}{$\delta_{\textrm{nc}}^{\textrm{ARIADNE}}$} & \multicolumn{1}{|c|}{$\delta_{\textrm{nc}}^{\textrm{LEPTO}}$} \\ \hline 
\endfirsthead

\multicolumn{7}{c}%
{ \tablename\ \thetable{} -- continued from previous page} \\
\hline
\endhead

\hline
\multicolumn{7}{c}%
{ \tablename\ \thetable{} -- continued on next page} \\
\hline
\endfoot

\endlastfoot 

1.5--2.5   &   $+$0.35   &   ${\pm2.6\times10^{-4}}$  &   $^{+2.0\times10^{-2}}_{-2.0\times10^{-2}}$   &   ${+2.8\times10^{-2}}$   &   ${+3.2\times10^{-2}}$   &   ${+2.3\times10^{-2}}$ \\
2.5--3.5   &   $+$0.29   &   ${\pm1.6\times10^{-4}}$  &   $^{+1.4\times10^{-2}}_{-1.4\times10^{-2}}$   &   ${+1.7\times10^{-2}}$   &   ${+1.9\times10^{-2}}$   &   ${+1.4\times10^{-2}}$ \\
3.5--4.5   &   $+$0.24   &   ${\pm1.2\times10^{-4}}$  &   $^{+7.2\times10^{-3}}_{-7.2\times10^{-3}}$   &   ${+4.7\times10^{-3}}$   &   ${+4.1\times10^{-3}}$   &   ${+5.2\times10^{-3}}$ \\
4.5--5.5   &   $+$0.19   &   ${\pm1.0\times10^{-4}}$  &   $^{+5.3\times10^{-3}}_{-5.5\times10^{-3}}$   &   ${+9.7\times10^{-4}}$   &   ${+1.3\times10^{-3}}$   &   ${+6.9\times10^{-4}}$ \\
5.5--6.5   &   $+$0.15   &   ${\pm9.2\times10^{-5}}$  &   $^{+4.2\times10^{-3}}_{-4.3\times10^{-3}}$   &   ${-2.3\times10^{-3}}$   &   ${-3.8\times10^{-3}}$   &   ${-8.1\times10^{-4}}$ \\
6.5--7.5   &   $+$0.12   &   ${\pm8.7\times10^{-5}}$  &   $^{+3.5\times10^{-3}}_{-3.7\times10^{-3}}$   &   ${-3.6\times10^{-3}}$   &   ${-3.6\times10^{-3}}$   &   ${-3.6\times10^{-3}}$ \\
7.5--8.5   &   $+$0.10   &   ${\pm8.4\times10^{-5}}$  &   $^{+2.9\times10^{-3}}_{-3.1\times10^{-3}}$   &   ${-4.6\times10^{-3}}$   &   ${-4.7\times10^{-3}}$   &   ${-4.6\times10^{-3}}$ \\
8.5--9.5   &   $+$0.08   &   ${\pm8.5\times10^{-5}}$  &   $^{+2.7\times10^{-3}}_{-2.8\times10^{-3}}$   &   ${-4.9\times10^{-3}}$   &   ${-4.7\times10^{-3}}$   &   ${-5.1\times10^{-3}}$ \\
\end{longtable}
\end{center}
\end{small}

%cn vs Nch

\clearpage
\begin{small}
\begin{center}
\renewcommand*{\arraystretch}{1.2}
\begin{longtable}{| c | c | c | c | c | c | c |}
\hline

\hline\multicolumn{1}{|c|}{$\Nch$} & \multicolumn{1}{|c|}{$c_1\{2\}$} & \multicolumn{1}{|c|}{$\delta_{\textrm{stat}}$} & \multicolumn{1}{|c|}{$\delta_{\textrm{syst}}$} & \multicolumn{1}{|c|}{$\delta_{\textrm{nc}}$} & \multicolumn{1}{|c|}{$\delta_{\textrm{nc}}^{\textrm{ARIADNE}}$} & \multicolumn{1}{|c|}{$\delta_{\textrm{nc}}^{\textrm{LEPTO}}$} \\ \hline 
\endfirsthead

\multicolumn{7}{c}%
{ \tablename\ \thetable{} -- continued from previous page} \\
\hline
\endhead

\hline
\multicolumn{7}{c}%
{ \tablename\ \thetable{} -- continued on next page} \\
\hline
\endfoot

\endlastfoot 

1.5--2.5   &   $+$0.023   &   ${\pm1.4\times10^{-3}}$  &   $^{+4.9\times10^{-3}}_{-4.8\times10^{-3}}$   &   ${+8.0\times10^{-3}}$   &   ${+8.6\times10^{-3}}$   &   ${+7.5\times10^{-3}}$ \\
2.5--3.5   &   $+$0.017   &   ${\pm6.9\times10^{-4}}$  &   $^{+2.9\times10^{-3}}_{-3.1\times10^{-3}}$   &   ${+5.7\times10^{-3}}$   &   ${+5.7\times10^{-3}}$   &   ${+5.6\times10^{-3}}$ \\
3.5--4.5   &   $+$0.009   &   ${\pm4.5\times10^{-4}}$  &   $^{+1.4\times10^{-3}}_{-1.7\times10^{-3}}$   &   ${+2.7\times10^{-3}}$   &   ${+2.2\times10^{-3}}$   &   ${+3.2\times10^{-3}}$ \\
4.5--5.5   &   $-$0.000   &   ${\pm3.4\times10^{-4}}$  &   $^{+1.3\times10^{-3}}_{-1.8\times10^{-3}}$   &   ${+1.8\times10^{-3}}$   &   ${+1.8\times10^{-3}}$   &   ${+1.7\times10^{-3}}$ \\
5.5--6.5   &   $-$0.009   &   ${\pm2.9\times10^{-4}}$  &   $^{+1.4\times10^{-3}}_{-1.8\times10^{-3}}$   &   ${+1.1\times10^{-3}}$   &   ${+1.2\times10^{-3}}$   &   ${+8.9\times10^{-4}}$ \\
6.5--7.5   &   $-$0.017   &   ${\pm2.6\times10^{-4}}$  &   $^{+1.5\times10^{-3}}_{-2.1\times10^{-3}}$   &   ${+1.6\times10^{-3}}$   &   ${+2.1\times10^{-3}}$   &   ${+1.2\times10^{-3}}$ \\
7.5--8.5   &   $-$0.027   &   ${\pm2.5\times10^{-4}}$  &   $^{+2.1\times10^{-3}}_{-2.5\times10^{-3}}$   &   ${+2.6\times10^{-3}}$   &   ${+3.1\times10^{-3}}$   &   ${+2.0\times10^{-3}}$ \\
8.5--9.5   &   $-$0.033   &   ${\pm2.4\times10^{-4}}$  &   $^{+2.6\times10^{-3}}_{-3.0\times10^{-3}}$   &   ${+3.4\times10^{-3}}$   &   ${+3.9\times10^{-3}}$   &   ${+2.9\times10^{-3}}$ \\
\end{longtable}
\end{center}
\end{small}

%cn vs Nch

\clearpage
\begin{small}
\begin{center}
\renewcommand*{\arraystretch}{1.2}
\begin{longtable}{| c | c | c | c | c | c | c |}
\hline

\hline\multicolumn{1}{|c|}{$\Nch$} & \multicolumn{1}{|c|}{$c_1\{2\}$} & \multicolumn{1}{|c|}{$\delta_{\textrm{stat}}$} & \multicolumn{1}{|c|}{$\delta_{\textrm{syst}}$} & \multicolumn{1}{|c|}{$\delta_{\textrm{nc}}$} & \multicolumn{1}{|c|}{$\delta_{\textrm{nc}}^{\textrm{ARIADNE}}$} & \multicolumn{1}{|c|}{$\delta_{\textrm{nc}}^{\textrm{LEPTO}}$} \\ \hline 
\endfirsthead

\multicolumn{7}{c}%
{ \tablename\ \thetable{} -- continued from previous page} \\
\hline
\endhead

\hline
\multicolumn{7}{c}%
{ \tablename\ \thetable{} -- continued on next page} \\
\hline
\endfoot

\endlastfoot 

1.5--2.5   &   $+$0.02   &   ${\pm3.3\times10^{-3}}$  &   $^{+1.0\times10^{-2}}_{-1.1\times10^{-2}}$   &   ${+1.3\times10^{-2}}$   &   ${+1.3\times10^{-2}}$   &   ${+1.3\times10^{-2}}$ \\
2.5--3.5   &   $+$0.00   &   ${\pm1.7\times10^{-3}}$  &   $^{+6.1\times10^{-3}}_{-6.4\times10^{-3}}$   &   ${+9.9\times10^{-3}}$   &   ${+9.5\times10^{-3}}$   &   ${+1.0\times10^{-2}}$ \\
3.5--4.5   &   $-$0.01   &   ${\pm1.1\times10^{-3}}$  &   $^{+3.6\times10^{-3}}_{-3.6\times10^{-3}}$   &   ${+5.9\times10^{-3}}$   &   ${+4.9\times10^{-3}}$   &   ${+7.0\times10^{-3}}$ \\
4.5--5.5   &   $-$0.03   &   ${\pm8.6\times10^{-4}}$  &   $^{+3.5\times10^{-3}}_{-3.5\times10^{-3}}$   &   ${+4.6\times10^{-3}}$   &   ${+4.5\times10^{-3}}$   &   ${+4.7\times10^{-3}}$ \\
5.5--6.5   &   $-$0.05   &   ${\pm7.3\times10^{-4}}$  &   $^{+3.4\times10^{-3}}_{-3.5\times10^{-3}}$   &   ${+3.9\times10^{-3}}$   &   ${+3.9\times10^{-3}}$   &   ${+3.9\times10^{-3}}$ \\
6.5--7.5   &   $-$0.07   &   ${\pm6.6\times10^{-4}}$  &   $^{+2.2\times10^{-3}}_{-2.6\times10^{-3}}$   &   ${+4.5\times10^{-3}}$   &   ${+4.5\times10^{-3}}$   &   ${+4.5\times10^{-3}}$ \\
7.5--8.5   &   $-$0.09   &   ${\pm6.3\times10^{-4}}$  &   $^{+2.3\times10^{-3}}_{-2.6\times10^{-3}}$   &   ${+5.1\times10^{-3}}$   &   ${+4.9\times10^{-3}}$   &   ${+5.3\times10^{-3}}$ \\
8.5--9.5   &   $-$0.10   &   ${\pm6.2\times10^{-4}}$  &   $^{+2.3\times10^{-3}}_{-2.6\times10^{-3}}$   &   ${+4.9\times10^{-3}}$   &   ${+4.1\times10^{-3}}$   &   ${+5.6\times10^{-3}}$ \\
\end{longtable}
\end{center}
\end{small}

%cn vs Nch

\clearpage
\begin{small}
\begin{center}
\renewcommand*{\arraystretch}{1.2}
\begin{longtable}{| c | c | c | c | c | c | c |}
\hline

\hline\multicolumn{1}{|c|}{$\Nch$} & \multicolumn{1}{|c|}{$c_2\{2\}$} & \multicolumn{1}{|c|}{$\delta_{\textrm{stat}}$} & \multicolumn{1}{|c|}{$\delta_{\textrm{syst}}$} & \multicolumn{1}{|c|}{$\delta_{\textrm{nc}}$} & \multicolumn{1}{|c|}{$\delta_{\textrm{nc}}^{\textrm{ARIADNE}}$} & \multicolumn{1}{|c|}{$\delta_{\textrm{nc}}^{\textrm{LEPTO}}$} \\ \hline 
\endfirsthead

\multicolumn{7}{c}%
{ \tablename\ \thetable{} -- continued from previous page} \\
\hline
\endhead

\hline
\multicolumn{7}{c}%
{ \tablename\ \thetable{} -- continued on next page} \\
\hline
\endfoot

\endlastfoot 

1.5--2.5   &   $+$0.146   &   ${\pm2.7\times10^{-4}}$  &   $^{+1.4\times10^{-2}}_{-1.4\times10^{-2}}$   &   ${+9.7\times10^{-3}}$   &   ${+1.2\times10^{-2}}$   &   ${+7.0\times10^{-3}}$ \\
2.5--3.5   &   $+$0.113   &   ${\pm1.7\times10^{-4}}$  &   $^{+9.2\times10^{-3}}_{-8.9\times10^{-3}}$   &   ${+5.2\times10^{-3}}$   &   ${+6.5\times10^{-3}}$   &   ${+3.9\times10^{-3}}$ \\
3.5--4.5   &   $+$0.090   &   ${\pm1.2\times10^{-4}}$  &   $^{+4.5\times10^{-3}}_{-3.8\times10^{-3}}$   &   ${+7.7\times10^{-4}}$   &   ${+6.6\times10^{-4}}$   &   ${+8.8\times10^{-4}}$ \\
4.5--5.5   &   $+$0.071   &   ${\pm1.0\times10^{-4}}$  &   $^{+3.1\times10^{-3}}_{-2.3\times10^{-3}}$   &   ${-7.6\times10^{-4}}$   &   ${-8.3\times10^{-4}}$   &   ${-6.8\times10^{-4}}$ \\
5.5--6.5   &   $+$0.058   &   ${\pm9.2\times10^{-5}}$  &   $^{+2.5\times10^{-3}}_{-1.7\times10^{-3}}$   &   ${-8.0\times10^{-4}}$   &   ${-1.0\times10^{-3}}$   &   ${-5.6\times10^{-4}}$ \\
6.5--7.5   &   $+$0.051   &   ${\pm8.6\times10^{-5}}$  &   $^{+2.1\times10^{-3}}_{-1.5\times10^{-3}}$   &   ${-1.6\times10^{-3}}$   &   ${-1.9\times10^{-3}}$   &   ${-1.3\times10^{-3}}$ \\
7.5--8.5   &   $+$0.048   &   ${\pm8.3\times10^{-5}}$  &   $^{+1.9\times10^{-3}}_{-1.5\times10^{-3}}$   &   ${-2.6\times10^{-3}}$   &   ${-2.9\times10^{-3}}$   &   ${-2.3\times10^{-3}}$ \\
8.5--9.5   &   $+$0.045   &   ${\pm8.3\times10^{-5}}$  &   $^{+1.7\times10^{-3}}_{-1.3\times10^{-3}}$   &   ${-3.3\times10^{-3}}$   &   ${-3.6\times10^{-3}}$   &   ${-3.1\times10^{-3}}$ \\
\end{longtable}
\end{center}
\end{small}

%cn vs Nch

\clearpage
\begin{small}
\begin{center}
\renewcommand*{\arraystretch}{1.2}
\begin{longtable}{| c | c | c | c | c | c | c |}
\hline

\hline\multicolumn{1}{|c|}{$\Nch$} & \multicolumn{1}{|c|}{$c_2\{2\}$} & \multicolumn{1}{|c|}{$\delta_{\textrm{stat}}$} & \multicolumn{1}{|c|}{$\delta_{\textrm{syst}}$} & \multicolumn{1}{|c|}{$\delta_{\textrm{nc}}$} & \multicolumn{1}{|c|}{$\delta_{\textrm{nc}}^{\textrm{ARIADNE}}$} & \multicolumn{1}{|c|}{$\delta_{\textrm{nc}}^{\textrm{LEPTO}}$} \\ \hline 
\endfirsthead

\multicolumn{7}{c}%
{ \tablename\ \thetable{} -- continued from previous page} \\
\hline
\endhead

\hline
\multicolumn{7}{c}%
{ \tablename\ \thetable{} -- continued on next page} \\
\hline
\endfoot

\endlastfoot 

1.5--2.5   &   $+$0.0120   &   ${\pm1.4\times10^{-3}}$  &   $^{+5.7\times10^{-3}}_{-5.7\times10^{-3}}$   &   ${-3.6\times10^{-3}}$   &   ${-4.0\times10^{-3}}$   &   ${-3.2\times10^{-3}}$ \\
2.5--3.5   &   $+$0.0034   &   ${\pm6.9\times10^{-4}}$  &   $^{+3.5\times10^{-3}}_{-3.5\times10^{-3}}$   &   ${+2.0\times10^{-4}}$   &   ${+2.5\times10^{-4}}$   &   ${+1.5\times10^{-4}}$ \\
3.5--4.5   &   $+$0.0006   &   ${\pm4.5\times10^{-4}}$  &   $^{+1.5\times10^{-3}}_{-1.5\times10^{-3}}$   &   ${+5.5\times10^{-4}}$   &   ${+6.8\times10^{-4}}$   &   ${+4.3\times10^{-4}}$ \\
4.5--5.5   &   $-$0.0007   &   ${\pm3.4\times10^{-4}}$  &   $^{+1.0\times10^{-3}}_{-9.7\times10^{-4}}$   &   ${+1.1\times10^{-3}}$   &   ${+1.2\times10^{-3}}$   &   ${+1.0\times10^{-3}}$ \\
5.5--6.5   &   $-$0.0004   &   ${\pm2.9\times10^{-4}}$  &   $^{+1.3\times10^{-3}}_{-1.2\times10^{-3}}$   &   ${+1.6\times10^{-3}}$   &   ${+1.6\times10^{-3}}$   &   ${+1.6\times10^{-3}}$ \\
6.5--7.5   &   $-$0.0004   &   ${\pm2.6\times10^{-4}}$  &   $^{+1.5\times10^{-3}}_{-1.4\times10^{-3}}$   &   ${+1.7\times10^{-3}}$   &   ${+1.7\times10^{-3}}$   &   ${+1.8\times10^{-3}}$ \\
7.5--8.5   &   $+$0.0007   &   ${\pm2.5\times10^{-4}}$  &   $^{+1.8\times10^{-3}}_{-1.7\times10^{-3}}$   &   ${+1.6\times10^{-3}}$   &   ${+1.6\times10^{-3}}$   &   ${+1.6\times10^{-3}}$ \\
8.5--9.5   &   $+$0.0014   &   ${\pm2.4\times10^{-4}}$  &   $^{+1.8\times10^{-3}}_{-1.7\times10^{-3}}$   &   ${+1.4\times10^{-3}}$   &   ${+1.5\times10^{-3}}$   &   ${+1.3\times10^{-3}}$ \\
\end{longtable}
\end{center}
\end{small}

%cn vs Nch

\clearpage
\begin{small}
\begin{center}
\renewcommand*{\arraystretch}{1.2}
\begin{longtable}{| c | c | c | c | c | c | c |}
\hline

\hline\multicolumn{1}{|c|}{$\Nch$} & \multicolumn{1}{|c|}{$c_2\{2\}$} & \multicolumn{1}{|c|}{$\delta_{\textrm{stat}}$} & \multicolumn{1}{|c|}{$\delta_{\textrm{syst}}$} & \multicolumn{1}{|c|}{$\delta_{\textrm{nc}}$} & \multicolumn{1}{|c|}{$\delta_{\textrm{nc}}^{\textrm{ARIADNE}}$} & \multicolumn{1}{|c|}{$\delta_{\textrm{nc}}^{\textrm{LEPTO}}$} \\ \hline 
\endfirsthead

\multicolumn{7}{c}%
{ \tablename\ \thetable{} -- continued from previous page} \\
\hline
\endhead

\hline
\multicolumn{7}{c}%
{ \tablename\ \thetable{} -- continued on next page} \\
\hline
\endfoot

\endlastfoot 

1.5--2.5   &   $+$0.013   &   ${\pm3.3\times10^{-3}}$  &   $^{+5.8\times10^{-3}}_{-6.0\times10^{-3}}$   &   ${-2.8\times10^{-3}}$   &   ${-1.8\times10^{-3}}$   &   ${-3.8\times10^{-3}}$ \\
2.5--3.5   &   $+$0.005   &   ${\pm1.7\times10^{-3}}$  &   $^{+3.4\times10^{-3}}_{-3.4\times10^{-3}}$   &   ${+6.4\times10^{-5}}$   &   ${-5.2\times10^{-4}}$   &   ${+6.4\times10^{-4}}$ \\
3.5--4.5   &   $+$0.003   &   ${\pm1.1\times10^{-3}}$  &   $^{+1.5\times10^{-3}}_{-1.4\times10^{-3}}$   &   ${+3.9\times10^{-4}}$   &   ${+1.8\times10^{-4}}$   &   ${+5.9\times10^{-4}}$ \\
4.5--5.5   &   $+$0.005   &   ${\pm8.6\times10^{-4}}$  &   $^{+9.2\times10^{-4}}_{-8.6\times10^{-4}}$   &   ${+1.1\times10^{-3}}$   &   ${+8.9\times10^{-4}}$   &   ${+1.3\times10^{-3}}$ \\
5.5--6.5   &   $+$0.005   &   ${\pm7.3\times10^{-4}}$  &   $^{+1.4\times10^{-3}}_{-1.4\times10^{-3}}$   &   ${+1.8\times10^{-3}}$   &   ${+1.9\times10^{-3}}$   &   ${+1.8\times10^{-3}}$ \\
6.5--7.5   &   $+$0.008   &   ${\pm6.6\times10^{-4}}$  &   $^{+1.8\times10^{-3}}_{-1.8\times10^{-3}}$   &   ${+2.2\times10^{-3}}$   &   ${+2.2\times10^{-3}}$   &   ${+2.1\times10^{-3}}$ \\
7.5--8.5   &   $+$0.011   &   ${\pm6.3\times10^{-4}}$  &   $^{+1.8\times10^{-3}}_{-1.8\times10^{-3}}$   &   ${+1.9\times10^{-3}}$   &   ${+2.3\times10^{-3}}$   &   ${+1.6\times10^{-3}}$ \\
8.5--9.5   &   $+$0.015   &   ${\pm6.2\times10^{-4}}$  &   $^{+1.8\times10^{-3}}_{-1.7\times10^{-3}}$   &   ${+1.8\times10^{-3}}$   &   ${+2.1\times10^{-3}}$   &   ${+1.5\times10^{-3}}$ \\
\end{longtable}
\end{center}
\end{small}

%cn vs Nch

\clearpage
\begin{small}
\begin{center}
\renewcommand*{\arraystretch}{1.2}
\begin{longtable}{| c | c | c | c | c | c | c |}
\hline

\hline\multicolumn{1}{|c|}{$\Nch$} & \multicolumn{1}{|c|}{$c_3\{2\}$} & \multicolumn{1}{|c|}{$\delta_{\textrm{stat}}$} & \multicolumn{1}{|c|}{$\delta_{\textrm{syst}}$} & \multicolumn{1}{|c|}{$\delta_{\textrm{nc}}$} & \multicolumn{1}{|c|}{$\delta_{\textrm{nc}}^{\textrm{ARIADNE}}$} & \multicolumn{1}{|c|}{$\delta_{\textrm{nc}}^{\textrm{LEPTO}}$} \\ \hline 
\endfirsthead

\multicolumn{7}{c}%
{ \tablename\ \thetable{} -- continued from previous page} \\
\hline
\endhead

\hline
\multicolumn{7}{c}%
{ \tablename\ \thetable{} -- continued on next page} \\
\hline
\endfoot

\endlastfoot 

1.5--2.5   &   $+$0.061   &   ${\pm2.7\times10^{-4}}$  &   $^{+5.5\times10^{-3}}_{-5.3\times10^{-3}}$   &   ${+3.9\times10^{-3}}$   &   ${+4.4\times10^{-3}}$   &   ${+3.4\times10^{-3}}$ \\
2.5--3.5   &   $+$0.046   &   ${\pm1.7\times10^{-4}}$  &   $^{+4.0\times10^{-3}}_{-3.8\times10^{-3}}$   &   ${+2.1\times10^{-3}}$   &   ${+2.3\times10^{-3}}$   &   ${+1.8\times10^{-3}}$ \\
3.5--4.5   &   $+$0.037   &   ${\pm1.2\times10^{-4}}$  &   $^{+2.5\times10^{-3}}_{-2.0\times10^{-3}}$   &   ${+2.2\times10^{-4}}$   &   ${+1.2\times10^{-4}}$   &   ${+3.1\times10^{-4}}$ \\
4.5--5.5   &   $+$0.029   &   ${\pm1.0\times10^{-4}}$  &   $^{+1.6\times10^{-3}}_{-1.0\times10^{-3}}$   &   ${-1.0\times10^{-3}}$   &   ${-1.2\times10^{-3}}$   &   ${-8.2\times10^{-4}}$ \\
5.5--6.5   &   $+$0.024   &   ${\pm9.2\times10^{-5}}$  &   $^{+1.2\times10^{-3}}_{-6.7\times10^{-4}}$   &   ${-9.3\times10^{-4}}$   &   ${-1.1\times10^{-3}}$   &   ${-7.7\times10^{-4}}$ \\
6.5--7.5   &   $+$0.021   &   ${\pm8.6\times10^{-5}}$  &   $^{+9.8\times10^{-4}}_{-5.8\times10^{-4}}$   &   ${-1.2\times10^{-3}}$   &   ${-1.4\times10^{-3}}$   &   ${-1.1\times10^{-3}}$ \\
7.5--8.5   &   $+$0.019   &   ${\pm8.3\times10^{-5}}$  &   $^{+8.1\times10^{-4}}_{-5.5\times10^{-4}}$   &   ${-1.6\times10^{-3}}$   &   ${-1.7\times10^{-3}}$   &   ${-1.4\times10^{-3}}$ \\
8.5--9.5   &   $+$0.017   &   ${\pm8.3\times10^{-5}}$  &   $^{+6.5\times10^{-4}}_{-5.1\times10^{-4}}$   &   ${-1.7\times10^{-3}}$   &   ${-1.9\times10^{-3}}$   &   ${-1.6\times10^{-3}}$ \\
\end{longtable}
\end{center}
\end{small}

%cn vs Nch

\clearpage
\begin{small}
\begin{center}
\renewcommand*{\arraystretch}{1.2}
\begin{longtable}{| c | c | c | c | c | c | c |}
\hline

\hline\multicolumn{1}{|c|}{$\Nch$} & \multicolumn{1}{|c|}{$c_3\{2\}$} & \multicolumn{1}{|c|}{$\delta_{\textrm{stat}}$} & \multicolumn{1}{|c|}{$\delta_{\textrm{syst}}$} & \multicolumn{1}{|c|}{$\delta_{\textrm{nc}}$} & \multicolumn{1}{|c|}{$\delta_{\textrm{nc}}^{\textrm{ARIADNE}}$} & \multicolumn{1}{|c|}{$\delta_{\textrm{nc}}^{\textrm{LEPTO}}$} \\ \hline 
\endfirsthead

\multicolumn{7}{c}%
{ \tablename\ \thetable{} -- continued from previous page} \\
\hline
\endhead

\hline
\multicolumn{7}{c}%
{ \tablename\ \thetable{} -- continued on next page} \\
\hline
\endfoot

\endlastfoot 

1.5--2.5   &   $+$0.0024   &   ${\pm1.4\times10^{-3}}$  &   $^{+8.7\times10^{-4}}_{-7.3\times10^{-4}}$   &   ${+5.7\times10^{-5}}$   &   ${+9.6\times10^{-5}}$   &   ${+1.8\times10^{-5}}$ \\
2.5--3.5   &   $-$0.0003   &   ${\pm6.9\times10^{-4}}$  &   $^{+5.8\times10^{-4}}_{-5.7\times10^{-4}}$   &   ${+5.7\times10^{-5}}$   &   ${+1.2\times10^{-4}}$   &   ${-3.2\times10^{-6}}$ \\
3.5--4.5   &   $+$0.0005   &   ${\pm4.5\times10^{-4}}$  &   $^{+7.6\times10^{-4}}_{-7.9\times10^{-4}}$   &   ${+7.6\times10^{-5}}$   &   ${+1.2\times10^{-4}}$   &   ${+3.0\times10^{-5}}$ \\
4.5--5.5   &   $-$0.0010   &   ${\pm3.4\times10^{-4}}$  &   $^{+8.8\times10^{-4}}_{-8.7\times10^{-4}}$   &   ${+6.9\times10^{-5}}$   &   ${+8.4\times10^{-6}}$   &   ${+1.3\times10^{-4}}$ \\
5.5--6.5   &   $-$0.0007   &   ${\pm2.9\times10^{-4}}$  &   $^{+8.3\times10^{-4}}_{-8.1\times10^{-4}}$   &   ${+2.2\times10^{-4}}$   &   ${+1.5\times10^{-4}}$   &   ${+3.0\times10^{-4}}$ \\
6.5--7.5   &   $-$0.0005   &   ${\pm2.6\times10^{-4}}$  &   $^{+3.8\times10^{-4}}_{-3.8\times10^{-4}}$   &   ${+3.1\times10^{-4}}$   &   ${+2.4\times10^{-4}}$   &   ${+3.8\times10^{-4}}$ \\
7.5--8.5   &   $-$0.0011   &   ${\pm2.5\times10^{-4}}$  &   $^{+3.0\times10^{-4}}_{-2.1\times10^{-4}}$   &   ${+3.9\times10^{-4}}$   &   ${+4.4\times10^{-4}}$   &   ${+3.3\times10^{-4}}$ \\
8.5--9.5   &   $-$0.0015   &   ${\pm2.4\times10^{-4}}$  &   $^{+4.5\times10^{-4}}_{-3.8\times10^{-4}}$   &   ${+3.9\times10^{-4}}$   &   ${+4.5\times10^{-4}}$   &   ${+3.2\times10^{-4}}$ \\
\end{longtable}
\end{center}
\end{small}

%cn vs Nch

\clearpage
\begin{small}
\begin{center}
\renewcommand*{\arraystretch}{1.2}
\begin{longtable}{| c | c | c | c | c | c | c |}
\hline

\hline\multicolumn{1}{|c|}{$\Nch$} & \multicolumn{1}{|c|}{$c_3\{2\}$} & \multicolumn{1}{|c|}{$\delta_{\textrm{stat}}$} & \multicolumn{1}{|c|}{$\delta_{\textrm{syst}}$} & \multicolumn{1}{|c|}{$\delta_{\textrm{nc}}$} & \multicolumn{1}{|c|}{$\delta_{\textrm{nc}}^{\textrm{ARIADNE}}$} & \multicolumn{1}{|c|}{$\delta_{\textrm{nc}}^{\textrm{LEPTO}}$} \\ \hline 
\endfirsthead

\multicolumn{7}{c}%
{ \tablename\ \thetable{} -- continued from previous page} \\
\hline
\endhead

\hline
\multicolumn{7}{c}%
{ \tablename\ \thetable{} -- continued on next page} \\
\hline
\endfoot

\endlastfoot 

1.5--2.5   &   $+$0.0045   &   ${\pm3.3\times10^{-3}}$  &   $^{+4.8\times10^{-3}}_{-4.2\times10^{-3}}$   &   ${+2.0\times10^{-4}}$   &   ${+7.0\times10^{-4}}$   &   ${-3.1\times10^{-4}}$ \\
2.5--3.5   &   $+$0.0001   &   ${\pm1.7\times10^{-3}}$  &   $^{+3.3\times10^{-3}}_{-3.2\times10^{-3}}$   &   ${+9.8\times10^{-5}}$   &   ${+3.5\times10^{-4}}$   &   ${-1.5\times10^{-4}}$ \\
3.5--4.5   &   $-$0.0006   &   ${\pm1.1\times10^{-3}}$  &   $^{+1.4\times10^{-3}}_{-1.6\times10^{-3}}$   &   ${+1.7\times10^{-4}}$   &   ${+2.2\times10^{-4}}$   &   ${+1.2\times10^{-4}}$ \\
4.5--5.5   &   $-$0.0004   &   ${\pm8.6\times10^{-4}}$  &   $^{+9.1\times10^{-4}}_{-1.0\times10^{-3}}$   &   ${+2.2\times10^{-4}}$   &   ${+3.2\times10^{-4}}$   &   ${+1.1\times10^{-4}}$ \\
5.5--6.5   &   $-$0.0029   &   ${\pm7.3\times10^{-4}}$  &   $^{+5.4\times10^{-4}}_{-5.4\times10^{-4}}$   &   ${+5.0\times10^{-4}}$   &   ${+3.0\times10^{-4}}$   &   ${+6.9\times10^{-4}}$ \\
6.5--7.5   &   $-$0.0027   &   ${\pm6.6\times10^{-4}}$  &   $^{+7.3\times10^{-4}}_{-6.0\times10^{-4}}$   &   ${+3.8\times10^{-4}}$   &   ${+1.6\times10^{-4}}$   &   ${+6.0\times10^{-4}}$ \\
7.5--8.5   &   $-$0.0034   &   ${\pm6.3\times10^{-4}}$  &   $^{+9.6\times10^{-4}}_{-8.1\times10^{-4}}$   &   ${+5.2\times10^{-4}}$   &   ${+4.2\times10^{-4}}$   &   ${+6.3\times10^{-4}}$ \\
8.5--9.5   &   $-$0.0039   &   ${\pm6.2\times10^{-4}}$  &   $^{+1.2\times10^{-3}}_{-1.1\times10^{-3}}$   &   ${+3.5\times10^{-4}}$   &   ${+4.4\times10^{-4}}$   &   ${+2.6\times10^{-4}}$ \\
\end{longtable}
\end{center}
\end{small}

%cn vs Nch

\clearpage
\begin{small}
\begin{center}
\renewcommand*{\arraystretch}{1.2}
\begin{longtable}{| c | c | c | c | c | c | c |}
\hline

\hline\multicolumn{1}{|c|}{$\Nch$} & \multicolumn{1}{|c|}{$c_4\{2\}$} & \multicolumn{1}{|c|}{$\delta_{\textrm{stat}}$} & \multicolumn{1}{|c|}{$\delta_{\textrm{syst}}$} & \multicolumn{1}{|c|}{$\delta_{\textrm{nc}}$} & \multicolumn{1}{|c|}{$\delta_{\textrm{nc}}^{\textrm{ARIADNE}}$} & \multicolumn{1}{|c|}{$\delta_{\textrm{nc}}^{\textrm{LEPTO}}$} \\ \hline 
\endfirsthead

\multicolumn{7}{c}%
{ \tablename\ \thetable{} -- continued from previous page} \\
\hline
\endhead

\hline
\multicolumn{7}{c}%
{ \tablename\ \thetable{} -- continued on next page} \\
\hline
\endfoot

\endlastfoot 

1.5--2.5   &   $+$0.025   &   ${\pm2.7\times10^{-4}}$  &   $^{+3.3\times10^{-3}}_{-3.2\times10^{-3}}$   &   ${+2.2\times10^{-3}}$   &   ${+2.0\times10^{-3}}$   &   ${+2.4\times10^{-3}}$ \\
2.5--3.5   &   $+$0.018   &   ${\pm1.7\times10^{-4}}$  &   $^{+2.4\times10^{-3}}_{-2.2\times10^{-3}}$   &   ${+1.3\times10^{-3}}$   &   ${+1.2\times10^{-3}}$   &   ${+1.4\times10^{-3}}$ \\
3.5--4.5   &   $+$0.015   &   ${\pm1.2\times10^{-4}}$  &   $^{+1.4\times10^{-3}}_{-1.0\times10^{-3}}$   &   ${+4.8\times10^{-4}}$   &   ${+4.4\times10^{-4}}$   &   ${+5.2\times10^{-4}}$ \\
4.5--5.5   &   $+$0.012   &   ${\pm1.0\times10^{-4}}$  &   $^{+9.7\times10^{-4}}_{-5.2\times10^{-4}}$   &   ${+2.6\times10^{-4}}$   &   ${+2.2\times10^{-4}}$   &   ${+2.9\times10^{-4}}$ \\
5.5--6.5   &   $+$0.010   &   ${\pm9.1\times10^{-5}}$  &   $^{+7.0\times10^{-4}}_{-3.3\times10^{-4}}$   &   ${+2.6\times10^{-4}}$   &   ${+2.3\times10^{-4}}$   &   ${+2.9\times10^{-4}}$ \\
6.5--7.5   &   $+$0.009   &   ${\pm8.6\times10^{-5}}$  &   $^{+5.4\times10^{-4}}_{-3.3\times10^{-4}}$   &   ${+1.8\times10^{-4}}$   &   ${+1.8\times10^{-4}}$   &   ${+1.8\times10^{-4}}$ \\
7.5--8.5   &   $+$0.009   &   ${\pm8.3\times10^{-5}}$  &   $^{+4.9\times10^{-4}}_{-3.4\times10^{-4}}$   &   ${-7.1\times10^{-4}}$   &   ${-1.1\times10^{-3}}$   &   ${-3.3\times10^{-4}}$ \\
8.5--9.5   &   $+$0.008   &   ${\pm8.3\times10^{-5}}$  &   $^{+4.2\times10^{-4}}_{-3.2\times10^{-4}}$   &   ${-1.0\times10^{-3}}$   &   ${-1.1\times10^{-3}}$   &   ${-1.0\times10^{-3}}$ \\
\end{longtable}
\end{center}
\end{small}

%cn vs Nch

\clearpage
\begin{small}
\begin{center}
\renewcommand*{\arraystretch}{1.2}
\begin{longtable}{| c | c | c | c | c | c | c |}
\hline

\hline\multicolumn{1}{|c|}{$\Nch$} & \multicolumn{1}{|c|}{$c_4\{2\}$} & \multicolumn{1}{|c|}{$\delta_{\textrm{stat}}$} & \multicolumn{1}{|c|}{$\delta_{\textrm{syst}}$} & \multicolumn{1}{|c|}{$\delta_{\textrm{nc}}$} & \multicolumn{1}{|c|}{$\delta_{\textrm{nc}}^{\textrm{ARIADNE}}$} & \multicolumn{1}{|c|}{$\delta_{\textrm{nc}}^{\textrm{LEPTO}}$} \\ \hline 
\endfirsthead

\multicolumn{7}{c}%
{ \tablename\ \thetable{} -- continued from previous page} \\
\hline
\endhead

\hline
\multicolumn{7}{c}%
{ \tablename\ \thetable{} -- continued on next page} \\
\hline
\endfoot

\endlastfoot 

1.5--2.5   &   $+$0.0022   &   ${\pm1.3\times10^{-3}}$  &   $^{+9.2\times10^{-4}}_{-5.3\times10^{-4}}$   &   ${-1.5\times10^{-4}}$   &   ${-5.7\times10^{-5}}$   &   ${-2.4\times10^{-4}}$ \\
2.5--3.5   &   $+$0.0003   &   ${\pm6.9\times10^{-4}}$  &   $^{+7.2\times10^{-4}}_{-6.1\times10^{-4}}$   &   ${+1.9\times10^{-4}}$   &   ${+9.3\times10^{-5}}$   &   ${+2.8\times10^{-4}}$ \\
3.5--4.5   &   $+$0.0005   &   ${\pm4.5\times10^{-4}}$  &   $^{+6.8\times10^{-4}}_{-6.6\times10^{-4}}$   &   ${+2.8\times10^{-4}}$   &   ${+2.0\times10^{-4}}$   &   ${+3.6\times10^{-4}}$ \\
4.5--5.5   &   $-$0.0009   &   ${\pm3.4\times10^{-4}}$  &   $^{+5.4\times10^{-4}}_{-5.2\times10^{-4}}$   &   ${+3.6\times10^{-4}}$   &   ${+2.9\times10^{-4}}$   &   ${+4.2\times10^{-4}}$ \\
5.5--6.5   &   $-$0.0006   &   ${\pm2.9\times10^{-4}}$  &   $^{+3.2\times10^{-4}}_{-2.9\times10^{-4}}$   &   ${+2.2\times10^{-4}}$   &   ${+1.9\times10^{-4}}$   &   ${+2.6\times10^{-4}}$ \\
6.5--7.5   &   $-$0.0007   &   ${\pm2.6\times10^{-4}}$  &   $^{+1.5\times10^{-4}}_{-1.5\times10^{-4}}$   &   ${+2.1\times10^{-4}}$   &   ${+2.0\times10^{-4}}$   &   ${+2.2\times10^{-4}}$ \\
7.5--8.5   &   $-$0.0013   &   ${\pm2.5\times10^{-4}}$  &   $^{+3.6\times10^{-4}}_{-3.5\times10^{-4}}$   &   ${+2.0\times10^{-4}}$   &   ${+1.7\times10^{-4}}$   &   ${+2.2\times10^{-4}}$ \\
8.5--9.5   &   $-$0.0009   &   ${\pm2.4\times10^{-4}}$  &   $^{+4.9\times10^{-4}}_{-4.7\times10^{-4}}$   &   ${+2.0\times10^{-4}}$   &   ${+2.8\times10^{-4}}$   &   ${+1.2\times10^{-4}}$ \\
\end{longtable}
\end{center}
\end{small}

%cn vs Nch

\clearpage
\begin{small}
\begin{center}
\renewcommand*{\arraystretch}{1.2}
\begin{longtable}{| c | c | c | c | c | c | c |}
\hline

\hline\multicolumn{1}{|c|}{$\Nch$} & \multicolumn{1}{|c|}{$c_4\{2\}$} & \multicolumn{1}{|c|}{$\delta_{\textrm{stat}}$} & \multicolumn{1}{|c|}{$\delta_{\textrm{syst}}$} & \multicolumn{1}{|c|}{$\delta_{\textrm{nc}}$} & \multicolumn{1}{|c|}{$\delta_{\textrm{nc}}^{\textrm{ARIADNE}}$} & \multicolumn{1}{|c|}{$\delta_{\textrm{nc}}^{\textrm{LEPTO}}$} \\ \hline 
\endfirsthead

\multicolumn{7}{c}%
{ \tablename\ \thetable{} -- continued from previous page} \\
\hline
\endhead

\hline
\multicolumn{7}{c}%
{ \tablename\ \thetable{} -- continued on next page} \\
\hline
\endfoot

\endlastfoot 

1.5--2.5   &   $+$0.0012   &   ${\pm3.3\times10^{-3}}$  &   $^{+3.3\times10^{-3}}_{-2.8\times10^{-3}}$   &   ${+2.0\times10^{-4}}$   &   ${-5.1\times10^{-4}}$   &   ${+9.1\times10^{-4}}$ \\
2.5--3.5   &   $-$0.0002   &   ${\pm1.7\times10^{-3}}$  &   $^{+2.1\times10^{-3}}_{-1.8\times10^{-3}}$   &   ${+2.0\times10^{-4}}$   &   ${-1.0\times10^{-4}}$   &   ${+5.1\times10^{-4}}$ \\
3.5--4.5   &   $-$0.0015   &   ${\pm1.1\times10^{-3}}$  &   $^{+7.8\times10^{-4}}_{-7.3\times10^{-4}}$   &   ${+4.4\times10^{-4}}$   &   ${+5.0\times10^{-4}}$   &   ${+3.7\times10^{-4}}$ \\
4.5--5.5   &   $-$0.0025   &   ${\pm8.6\times10^{-4}}$  &   $^{+8.8\times10^{-4}}_{-8.1\times10^{-4}}$   &   ${+7.2\times10^{-4}}$   &   ${+7.7\times10^{-4}}$   &   ${+6.7\times10^{-4}}$ \\
5.5--6.5   &   $-$0.0012   &   ${\pm7.3\times10^{-4}}$  &   $^{+8.0\times10^{-4}}_{-7.3\times10^{-4}}$   &   ${+9.0\times10^{-4}}$   &   ${+7.6\times10^{-4}}$   &   ${+1.0\times10^{-3}}$ \\
6.5--7.5   &   $-$0.0026   &   ${\pm6.6\times10^{-4}}$  &   $^{+5.8\times10^{-4}}_{-5.3\times10^{-4}}$   &   ${+8.7\times10^{-4}}$   &   ${+9.3\times10^{-4}}$   &   ${+8.1\times10^{-4}}$ \\
7.5--8.5   &   $-$0.0026   &   ${\pm6.3\times10^{-4}}$  &   $^{+5.3\times10^{-4}}_{-5.4\times10^{-4}}$   &   ${+9.8\times10^{-4}}$   &   ${+1.2\times10^{-3}}$   &   ${+7.5\times10^{-4}}$ \\
8.5--9.5   &   $-$0.0020   &   ${\pm6.2\times10^{-4}}$  &   $^{+4.1\times10^{-4}}_{-4.1\times10^{-4}}$   &   ${+1.1\times10^{-3}}$   &   ${+1.3\times10^{-3}}$   &   ${+9.1\times10^{-4}}$ \\
\end{longtable}
\end{center}
\end{small}

%cn vs deta

\clearpage
\begin{small}
\begin{center}
\renewcommand*{\arraystretch}{1.2}
\begin{longtable}{| c | c | c | c | c | c | c |}
\hline

\hline\multicolumn{1}{|c|}{$\deta$} & \multicolumn{1}{|c|}{$c_1\{2\}$} & \multicolumn{1}{|c|}{$\delta_{\textrm{stat}}$} & \multicolumn{1}{|c|}{$\delta_{\textrm{syst}}$} & \multicolumn{1}{|c|}{$\delta_{\textrm{nc}}$} & \multicolumn{1}{|c|}{$\delta_{\textrm{nc}}^{\textrm{ARIADNE}}$} & \multicolumn{1}{|c|}{$\delta_{\textrm{nc}}^{\textrm{LEPTO}}$} \\ \hline 
\endfirsthead

\multicolumn{7}{c}%
{ \tablename\ \thetable{} -- continued from previous page} \\
\hline
\endhead

\hline
\multicolumn{7}{c}%
{ \tablename\ \thetable{} -- continued on next page} \\
\hline
\endfoot

\endlastfoot 

0--0.2   &   $+$0.14   &   ${\pm1.9\times10^{-4}}$  &   $^{+3.6\times10^{-3}}_{-4.1\times10^{-3}}$   &   ${-1.1\times10^{-2}}$   &   ${-1.1\times10^{-2}}$   &   ${-1.1\times10^{-2}}$ \\
0.2--0.4   &   $+$0.10   &   ${\pm2.0\times10^{-4}}$  &   $^{+2.4\times10^{-3}}_{-3.0\times10^{-3}}$   &   ${-7.2\times10^{-3}}$   &   ${-7.0\times10^{-3}}$   &   ${-7.3\times10^{-3}}$ \\
0.4--0.6   &   $+$0.07   &   ${\pm2.1\times10^{-4}}$  &   $^{+1.4\times10^{-3}}_{-2.1\times10^{-3}}$   &   ${-2.9\times10^{-3}}$   &   ${-2.7\times10^{-3}}$   &   ${-3.1\times10^{-3}}$ \\
0.6--0.8   &   $+$0.04   &   ${\pm2.2\times10^{-4}}$  &   $^{+1.1\times10^{-3}}_{-1.7\times10^{-3}}$   &   ${+7.0\times10^{-4}}$   &   ${+8.4\times10^{-4}}$   &   ${+5.5\times10^{-4}}$ \\
0.8--1   &   $+$0.01   &   ${\pm2.4\times10^{-4}}$  &   $^{+1.6\times10^{-3}}_{-1.8\times10^{-3}}$   &   ${+9.0\times10^{-4}}$   &   ${+1.2\times10^{-3}}$   &   ${+5.7\times10^{-4}}$ \\
1--1.2   &   $-$0.01   &   ${\pm2.5\times10^{-4}}$  &   $^{+2.4\times10^{-3}}_{-2.4\times10^{-3}}$   &   ${+1.7\times10^{-3}}$   &   ${+2.0\times10^{-3}}$   &   ${+1.3\times10^{-3}}$ \\
1.2--1.4   &   $-$0.02   &   ${\pm2.7\times10^{-4}}$  &   $^{+2.7\times10^{-3}}_{-2.7\times10^{-3}}$   &   ${+2.3\times10^{-3}}$   &   ${+2.6\times10^{-3}}$   &   ${+2.0\times10^{-3}}$ \\
1.4--1.6   &   $-$0.04   &   ${\pm2.9\times10^{-4}}$  &   $^{+2.9\times10^{-3}}_{-2.9\times10^{-3}}$   &   ${+3.7\times10^{-3}}$   &   ${+3.9\times10^{-3}}$   &   ${+3.6\times10^{-3}}$ \\
\end{longtable}
\end{center}
\end{small}

%cn vs deta

\clearpage
\begin{small}
\begin{center}
\renewcommand*{\arraystretch}{1.2}
\begin{longtable}{| c | c | c | c | c | c | c |}
\hline

\hline\multicolumn{1}{|c|}{$\deta$} & \multicolumn{1}{|c|}{$c_1\{2\}$} & \multicolumn{1}{|c|}{$\delta_{\textrm{stat}}$} & \multicolumn{1}{|c|}{$\delta_{\textrm{syst}}$} & \multicolumn{1}{|c|}{$\delta_{\textrm{nc}}$} & \multicolumn{1}{|c|}{$\delta_{\textrm{nc}}^{\textrm{ARIADNE}}$} & \multicolumn{1}{|c|}{$\delta_{\textrm{nc}}^{\textrm{LEPTO}}$} \\ \hline 
\endfirsthead

\multicolumn{7}{c}%
{ \tablename\ \thetable{} -- continued from previous page} \\
\hline
\endhead

\hline
\multicolumn{7}{c}%
{ \tablename\ \thetable{} -- continued on next page} \\
\hline
\endfoot

\endlastfoot 

0--0.2   &   $+$0.28   &   ${\pm4.3\times10^{-4}}$  &   $^{+2.5\times10^{-3}}_{-3.8\times10^{-3}}$   &   ${+1.1\times10^{-2}}$   &   ${+1.0\times10^{-2}}$   &   ${+1.2\times10^{-2}}$ \\
0.2--0.4   &   $+$0.21   &   ${\pm4.6\times10^{-4}}$  &   $^{+3.3\times10^{-3}}_{-4.2\times10^{-3}}$   &   ${+1.1\times10^{-2}}$   &   ${+9.7\times10^{-3}}$   &   ${+1.2\times10^{-2}}$ \\
0.4--0.6   &   $+$0.12   &   ${\pm5.0\times10^{-4}}$  &   $^{+3.9\times10^{-3}}_{-4.5\times10^{-3}}$   &   ${+9.7\times10^{-3}}$   &   ${+8.2\times10^{-3}}$   &   ${+1.1\times10^{-2}}$ \\
0.6--0.8   &   $+$0.04   &   ${\pm5.4\times10^{-4}}$  &   $^{+4.0\times10^{-3}}_{-4.2\times10^{-3}}$   &   ${+9.0\times10^{-3}}$   &   ${+6.8\times10^{-3}}$   &   ${+1.1\times10^{-2}}$ \\
0.8--1   &   $-$0.03   &   ${\pm5.8\times10^{-4}}$  &   $^{+3.8\times10^{-3}}_{-3.9\times10^{-3}}$   &   ${+7.2\times10^{-3}}$   &   ${+4.6\times10^{-3}}$   &   ${+9.9\times10^{-3}}$ \\
1--1.2   &   $-$0.08   &   ${\pm6.2\times10^{-4}}$  &   $^{+3.9\times10^{-3}}_{-3.9\times10^{-3}}$   &   ${+5.8\times10^{-3}}$   &   ${+2.9\times10^{-3}}$   &   ${+8.8\times10^{-3}}$ \\
1.2--1.4   &   $-$0.11   &   ${\pm6.6\times10^{-4}}$  &   $^{+3.7\times10^{-3}}_{-3.7\times10^{-3}}$   &   ${+3.9\times10^{-3}}$   &   ${+8.1\times10^{-4}}$   &   ${+7.0\times10^{-3}}$ \\
1.4--1.6   &   $-$0.13   &   ${\pm7.0\times10^{-4}}$  &   $^{+3.4\times10^{-3}}_{-3.5\times10^{-3}}$   &   ${+4.5\times10^{-3}}$   &   ${+1.3\times10^{-3}}$   &   ${+7.7\times10^{-3}}$ \\
\end{longtable}
\end{center}
\end{small}

%cn vs deta

\clearpage
\begin{small}
\begin{center}
\renewcommand*{\arraystretch}{1.2}
\begin{longtable}{| c | c | c | c | c | c | c |}
\hline

\hline\multicolumn{1}{|c|}{$\deta$} & \multicolumn{1}{|c|}{$c_2\{2\}$} & \multicolumn{1}{|c|}{$\delta_{\textrm{stat}}$} & \multicolumn{1}{|c|}{$\delta_{\textrm{syst}}$} & \multicolumn{1}{|c|}{$\delta_{\textrm{nc}}$} & \multicolumn{1}{|c|}{$\delta_{\textrm{nc}}^{\textrm{ARIADNE}}$} & \multicolumn{1}{|c|}{$\delta_{\textrm{nc}}^{\textrm{LEPTO}}$} \\ \hline 
\endfirsthead

\multicolumn{7}{c}%
{ \tablename\ \thetable{} -- continued from previous page} \\
\hline
\endhead

\hline
\multicolumn{7}{c}%
{ \tablename\ \thetable{} -- continued on next page} \\
\hline
\endfoot

\endlastfoot 

0--0.2   &   $+$0.101   &   ${\pm1.9\times10^{-4}}$  &   $^{+4.0\times10^{-3}}_{-3.9\times10^{-3}}$   &   ${-1.0\times10^{-2}}$   &   ${-1.0\times10^{-2}}$   &   ${-9.8\times10^{-3}}$ \\
0.2--0.4   &   $+$0.076   &   ${\pm2.0\times10^{-4}}$  &   $^{+3.2\times10^{-3}}_{-3.1\times10^{-3}}$   &   ${-6.7\times10^{-3}}$   &   ${-7.1\times10^{-3}}$   &   ${-6.4\times10^{-3}}$ \\
0.4--0.6   &   $+$0.057   &   ${\pm2.1\times10^{-4}}$  &   $^{+2.2\times10^{-3}}_{-2.1\times10^{-3}}$   &   ${-3.3\times10^{-3}}$   &   ${-3.7\times10^{-3}}$   &   ${-2.9\times10^{-3}}$ \\
0.6--0.8   &   $+$0.043   &   ${\pm2.2\times10^{-4}}$  &   $^{+1.6\times10^{-3}}_{-1.4\times10^{-3}}$   &   ${-2.5\times10^{-3}}$   &   ${-2.9\times10^{-3}}$   &   ${-2.0\times10^{-3}}$ \\
0.8--1   &   $+$0.031   &   ${\pm2.3\times10^{-4}}$  &   $^{+1.0\times10^{-3}}_{-8.1\times10^{-4}}$   &   ${-2.1\times10^{-3}}$   &   ${-2.5\times10^{-3}}$   &   ${-1.7\times10^{-3}}$ \\
1--1.2   &   $+$0.024   &   ${\pm2.5\times10^{-4}}$  &   $^{+6.2\times10^{-4}}_{-4.8\times10^{-4}}$   &   ${-1.6\times10^{-3}}$   &   ${-2.1\times10^{-3}}$   &   ${-1.1\times10^{-3}}$ \\
1.2--1.4   &   $+$0.019   &   ${\pm2.7\times10^{-4}}$  &   $^{+5.2\times10^{-4}}_{-3.6\times10^{-4}}$   &   ${-1.3\times10^{-3}}$   &   ${-1.8\times10^{-3}}$   &   ${-8.7\times10^{-4}}$ \\
1.4--1.6   &   $+$0.015   &   ${\pm2.9\times10^{-4}}$  &   $^{+5.1\times10^{-4}}_{-3.4\times10^{-4}}$   &   ${-7.3\times10^{-4}}$   &   ${-1.1\times10^{-3}}$   &   ${-3.5\times10^{-4}}$ \\
\end{longtable}
\end{center}
\end{small}

%cn vs deta

\clearpage
\begin{small}
\begin{center}
\renewcommand*{\arraystretch}{1.2}
\begin{longtable}{| c | c | c | c | c | c | c |}
\hline

\hline\multicolumn{1}{|c|}{$\deta$} & \multicolumn{1}{|c|}{$c_2\{2\}$} & \multicolumn{1}{|c|}{$\delta_{\textrm{stat}}$} & \multicolumn{1}{|c|}{$\delta_{\textrm{syst}}$} & \multicolumn{1}{|c|}{$\delta_{\textrm{nc}}$} & \multicolumn{1}{|c|}{$\delta_{\textrm{nc}}^{\textrm{ARIADNE}}$} & \multicolumn{1}{|c|}{$\delta_{\textrm{nc}}^{\textrm{LEPTO}}$} \\ \hline 
\endfirsthead

\multicolumn{7}{c}%
{ \tablename\ \thetable{} -- continued from previous page} \\
\hline
\endhead

\hline
\multicolumn{7}{c}%
{ \tablename\ \thetable{} -- continued on next page} \\
\hline
\endfoot

\endlastfoot 

0--0.2   &   $+$0.273   &   ${\pm3.9\times10^{-4}}$  &   $^{+2.3\times10^{-3}}_{-2.3\times10^{-3}}$   &   ${+1.0\times10^{-2}}$   &   ${+1.2\times10^{-2}}$   &   ${+8.5\times10^{-3}}$ \\
0.2--0.4   &   $+$0.228   &   ${\pm4.2\times10^{-4}}$  &   $^{+1.8\times10^{-3}}_{-2.1\times10^{-3}}$   &   ${+9.2\times10^{-3}}$   &   ${+1.1\times10^{-2}}$   &   ${+7.4\times10^{-3}}$ \\
0.4--0.6   &   $+$0.175   &   ${\pm4.6\times10^{-4}}$  &   $^{+1.4\times10^{-3}}_{-1.7\times10^{-3}}$   &   ${+7.6\times10^{-3}}$   &   ${+9.4\times10^{-3}}$   &   ${+5.8\times10^{-3}}$ \\
0.6--0.8   &   $+$0.128   &   ${\pm5.1\times10^{-4}}$  &   $^{+1.2\times10^{-3}}_{-1.2\times10^{-3}}$   &   ${+6.7\times10^{-3}}$   &   ${+8.4\times10^{-3}}$   &   ${+5.0\times10^{-3}}$ \\
0.8--1   &   $+$0.092   &   ${\pm5.5\times10^{-4}}$  &   $^{+8.6\times10^{-4}}_{-6.2\times10^{-4}}$   &   ${+6.0\times10^{-3}}$   &   ${+7.7\times10^{-3}}$   &   ${+4.3\times10^{-3}}$ \\
1--1.2   &   $+$0.071   &   ${\pm6.0\times10^{-4}}$  &   $^{+6.4\times10^{-4}}_{-3.0\times10^{-4}}$   &   ${+5.7\times10^{-3}}$   &   ${+7.4\times10^{-3}}$   &   ${+3.9\times10^{-3}}$ \\
1.2--1.4   &   $+$0.056   &   ${\pm6.5\times10^{-4}}$  &   $^{+7.2\times10^{-4}}_{-4.5\times10^{-4}}$   &   ${+5.0\times10^{-3}}$   &   ${+6.8\times10^{-3}}$   &   ${+3.1\times10^{-3}}$ \\
1.4--1.6   &   $+$0.046   &   ${\pm7.0\times10^{-4}}$  &   $^{+9.9\times10^{-4}}_{-7.9\times10^{-4}}$   &   ${+4.9\times10^{-3}}$   &   ${+6.6\times10^{-3}}$   &   ${+3.2\times10^{-3}}$ \\
\end{longtable}
\end{center}
\end{small}

%cn vs deta

\clearpage
\begin{small}
\begin{center}
\renewcommand*{\arraystretch}{1.2}
\begin{longtable}{| c | c | c | c | c | c | c |}
\hline

\hline\multicolumn{1}{|c|}{$\deta$} & \multicolumn{1}{|c|}{$c_3\{2\}$} & \multicolumn{1}{|c|}{$\delta_{\textrm{stat}}$} & \multicolumn{1}{|c|}{$\delta_{\textrm{syst}}$} & \multicolumn{1}{|c|}{$\delta_{\textrm{nc}}$} & \multicolumn{1}{|c|}{$\delta_{\textrm{nc}}^{\textrm{ARIADNE}}$} & \multicolumn{1}{|c|}{$\delta_{\textrm{nc}}^{\textrm{LEPTO}}$} \\ \hline 
\endfirsthead

\multicolumn{7}{c}%
{ \tablename\ \thetable{} -- continued from previous page} \\
\hline
\endhead

\hline
\multicolumn{7}{c}%
{ \tablename\ \thetable{} -- continued on next page} \\
\hline
\endfoot

\endlastfoot 

0--0.2   &   $+$0.0477   &   ${\pm1.9\times10^{-4}}$  &   $^{+2.1\times10^{-3}}_{-1.9\times10^{-3}}$   &   ${-5.3\times10^{-3}}$   &   ${-5.4\times10^{-3}}$   &   ${-5.2\times10^{-3}}$ \\
0.2--0.4   &   $+$0.0269   &   ${\pm2.0\times10^{-4}}$  &   $^{+1.3\times10^{-3}}_{-1.2\times10^{-3}}$   &   ${-2.7\times10^{-3}}$   &   ${-2.7\times10^{-3}}$   &   ${-2.6\times10^{-3}}$ \\
0.4--0.6   &   $+$0.0155   &   ${\pm2.1\times10^{-4}}$  &   $^{+5.1\times10^{-4}}_{-5.1\times10^{-4}}$   &   ${-7.2\times10^{-5}}$   &   ${-8.9\times10^{-6}}$   &   ${-1.3\times10^{-4}}$ \\
0.6--0.8   &   $+$0.0071   &   ${\pm2.2\times10^{-4}}$  &   $^{+2.5\times10^{-4}}_{-2.7\times10^{-4}}$   &   ${+1.2\times10^{-5}}$   &   ${+1.2\times10^{-4}}$   &   ${-9.4\times10^{-5}}$ \\
0.8--1   &   $+$0.0015   &   ${\pm2.3\times10^{-4}}$  &   $^{+3.3\times10^{-4}}_{-3.7\times10^{-4}}$   &   ${+1.2\times10^{-4}}$   &   ${+2.1\times10^{-4}}$   &   ${+3.6\times10^{-5}}$ \\
1--1.2   &   $-$0.0009   &   ${\pm2.5\times10^{-4}}$  &   $^{+5.7\times10^{-4}}_{-5.8\times10^{-4}}$   &   ${+2.8\times10^{-4}}$   &   ${+3.3\times10^{-4}}$   &   ${+2.2\times10^{-4}}$ \\
1.2--1.4   &   $-$0.0025   &   ${\pm2.7\times10^{-4}}$  &   $^{+5.9\times10^{-4}}_{-5.9\times10^{-4}}$   &   ${+3.8\times10^{-4}}$   &   ${+4.1\times10^{-4}}$   &   ${+3.4\times10^{-4}}$ \\
1.4--1.6   &   $-$0.0029   &   ${\pm2.9\times10^{-4}}$  &   $^{+5.3\times10^{-4}}_{-5.0\times10^{-4}}$   &   ${+3.5\times10^{-4}}$   &   ${+2.9\times10^{-4}}$   &   ${+4.2\times10^{-4}}$ \\
\end{longtable}
\end{center}
\end{small}

%cn vs deta

\clearpage
\begin{small}
\begin{center}
\renewcommand*{\arraystretch}{1.2}
\begin{longtable}{| c | c | c | c | c | c | c |}
\hline

\hline\multicolumn{1}{|c|}{$\deta$} & \multicolumn{1}{|c|}{$c_3\{2\}$} & \multicolumn{1}{|c|}{$\delta_{\textrm{stat}}$} & \multicolumn{1}{|c|}{$\delta_{\textrm{syst}}$} & \multicolumn{1}{|c|}{$\delta_{\textrm{nc}}$} & \multicolumn{1}{|c|}{$\delta_{\textrm{nc}}^{\textrm{ARIADNE}}$} & \multicolumn{1}{|c|}{$\delta_{\textrm{nc}}^{\textrm{LEPTO}}$} \\ \hline 
\endfirsthead

\multicolumn{7}{c}%
{ \tablename\ \thetable{} -- continued from previous page} \\
\hline
\endhead

\hline
\multicolumn{7}{c}%
{ \tablename\ \thetable{} -- continued on next page} \\
\hline
\endfoot

\endlastfoot 

0--0.2   &   $+$0.137   &   ${\pm4.0\times10^{-4}}$  &   $^{+1.7\times10^{-3}}_{-1.8\times10^{-3}}$   &   ${+7.6\times10^{-3}}$   &   ${+7.6\times10^{-3}}$   &   ${+7.7\times10^{-3}}$ \\
0.2--0.4   &   $+$0.097   &   ${\pm4.3\times10^{-4}}$  &   $^{+1.2\times10^{-3}}_{-1.4\times10^{-3}}$   &   ${+5.8\times10^{-3}}$   &   ${+5.8\times10^{-3}}$   &   ${+5.9\times10^{-3}}$ \\
0.4--0.6   &   $+$0.057   &   ${\pm4.7\times10^{-4}}$  &   $^{+7.2\times10^{-4}}_{-1.3\times10^{-3}}$   &   ${+3.3\times10^{-3}}$   &   ${+3.1\times10^{-3}}$   &   ${+3.6\times10^{-3}}$ \\
0.6--0.8   &   $+$0.026   &   ${\pm5.1\times10^{-4}}$  &   $^{+7.0\times10^{-4}}_{-1.1\times10^{-3}}$   &   ${+1.7\times10^{-3}}$   &   ${+1.2\times10^{-3}}$   &   ${+2.2\times10^{-3}}$ \\
0.8--1   &   $+$0.006   &   ${\pm5.6\times10^{-4}}$  &   $^{+9.3\times10^{-4}}_{-1.2\times10^{-3}}$   &   ${+8.3\times10^{-4}}$   &   ${+1.2\times10^{-4}}$   &   ${+1.6\times10^{-3}}$ \\
1--1.2   &   $-$0.004   &   ${\pm6.0\times10^{-4}}$  &   $^{+6.7\times10^{-4}}_{-9.1\times10^{-4}}$   &   ${+6.4\times10^{-4}}$   &   ${-1.5\times10^{-4}}$   &   ${+1.4\times10^{-3}}$ \\
1.2--1.4   &   $-$0.009   &   ${\pm6.5\times10^{-4}}$  &   $^{+5.9\times10^{-4}}_{-1.1\times10^{-3}}$   &   ${+1.8\times10^{-4}}$   &   ${-4.4\times10^{-4}}$   &   ${+8.0\times10^{-4}}$ \\
1.4--1.6   &   $-$0.009   &   ${\pm7.0\times10^{-4}}$  &   $^{+4.0\times10^{-4}}_{-7.8\times10^{-4}}$   &   ${+5.1\times10^{-5}}$   &   ${-3.2\times10^{-4}}$   &   ${+4.2\times10^{-4}}$ \\
\end{longtable}
\end{center}
\end{small}

%cn vs deta

\clearpage
\begin{small}
\begin{center}
\renewcommand*{\arraystretch}{1.2}
\begin{longtable}{| c | c | c | c | c | c | c |}
\hline

\hline\multicolumn{1}{|c|}{$\deta$} & \multicolumn{1}{|c|}{$c_4\{2\}$} & \multicolumn{1}{|c|}{$\delta_{\textrm{stat}}$} & \multicolumn{1}{|c|}{$\delta_{\textrm{syst}}$} & \multicolumn{1}{|c|}{$\delta_{\textrm{nc}}$} & \multicolumn{1}{|c|}{$\delta_{\textrm{nc}}^{\textrm{ARIADNE}}$} & \multicolumn{1}{|c|}{$\delta_{\textrm{nc}}^{\textrm{LEPTO}}$} \\ \hline 
\endfirsthead

\multicolumn{7}{c}%
{ \tablename\ \thetable{} -- continued from previous page} \\
\hline
\endhead

\hline
\multicolumn{7}{c}%
{ \tablename\ \thetable{} -- continued on next page} \\
\hline
\endfoot

\endlastfoot 

0--0.2   &   $+$0.031   &   ${\pm1.9\times10^{-4}}$  &   $^{+1.1\times10^{-3}}_{-1.2\times10^{-3}}$   &   ${+2.4\times10^{-4}}$   &   ${+2.0\times10^{-4}}$   &   ${+2.8\times10^{-4}}$ \\
0.2--0.4   &   $+$0.014   &   ${\pm2.0\times10^{-4}}$  &   $^{+5.9\times10^{-4}}_{-6.9\times10^{-4}}$   &   ${+4.7\times10^{-4}}$   &   ${+4.0\times10^{-4}}$   &   ${+5.3\times10^{-4}}$ \\
0.4--0.6   &   $+$0.008   &   ${\pm2.1\times10^{-4}}$  &   $^{+2.4\times10^{-4}}_{-2.8\times10^{-4}}$   &   ${+5.1\times10^{-4}}$   &   ${+4.8\times10^{-4}}$   &   ${+5.4\times10^{-4}}$ \\
0.6--0.8   &   $+$0.004   &   ${\pm2.2\times10^{-4}}$  &   $^{+1.3\times10^{-4}}_{-1.9\times10^{-4}}$   &   ${+2.9\times10^{-4}}$   &   ${+3.1\times10^{-4}}$   &   ${+2.8\times10^{-4}}$ \\
0.8--1   &   $+$0.002   &   ${\pm2.3\times10^{-4}}$  &   $^{+2.2\times10^{-4}}_{-2.5\times10^{-4}}$   &   ${+1.3\times10^{-4}}$   &   ${+2.1\times10^{-4}}$   &   ${+5.2\times10^{-5}}$ \\
1--1.2   &   $+$0.000   &   ${\pm2.5\times10^{-4}}$  &   $^{+3.3\times10^{-4}}_{-3.7\times10^{-4}}$   &   ${+1.2\times10^{-4}}$   &   ${+2.1\times10^{-4}}$   &   ${+2.4\times10^{-5}}$ \\
1.2--1.4   &   $+$0.001   &   ${\pm2.7\times10^{-4}}$  &   $^{+2.3\times10^{-4}}_{-2.9\times10^{-4}}$   &   ${+9.2\times10^{-5}}$   &   ${+1.8\times10^{-4}}$   &   ${+6.2\times10^{-6}}$ \\
1.4--1.6   &   $+$0.000   &   ${\pm2.9\times10^{-4}}$  &   $^{+2.8\times10^{-4}}_{-3.4\times10^{-4}}$   &   ${+3.1\times10^{-5}}$   &   ${+7.6\times10^{-5}}$   &   ${-1.5\times10^{-5}}$ \\
\end{longtable}
\end{center}
\end{small}

%cn vs deta

\clearpage
\begin{small}
\begin{center}
\renewcommand*{\arraystretch}{1.2}
\begin{longtable}{| c | c | c | c | c | c | c |}
\hline

\hline\multicolumn{1}{|c|}{$\deta$} & \multicolumn{1}{|c|}{$c_4\{2\}$} & \multicolumn{1}{|c|}{$\delta_{\textrm{stat}}$} & \multicolumn{1}{|c|}{$\delta_{\textrm{syst}}$} & \multicolumn{1}{|c|}{$\delta_{\textrm{nc}}$} & \multicolumn{1}{|c|}{$\delta_{\textrm{nc}}^{\textrm{ARIADNE}}$} & \multicolumn{1}{|c|}{$\delta_{\textrm{nc}}^{\textrm{LEPTO}}$} \\ \hline 
\endfirsthead

\multicolumn{7}{c}%
{ \tablename\ \thetable{} -- continued from previous page} \\
\hline
\endhead

\hline
\multicolumn{7}{c}%
{ \tablename\ \thetable{} -- continued on next page} \\
\hline
\endfoot

\endlastfoot 

0--0.2   &   $+$0.083   &   ${\pm4.0\times10^{-4}}$  &   $^{+1.1\times10^{-3}}_{-9.3\times10^{-4}}$   &   ${+1.0\times10^{-2}}$   &   ${+1.1\times10^{-2}}$   &   ${+9.6\times10^{-3}}$ \\
0.2--0.4   &   $+$0.054   &   ${\pm4.3\times10^{-4}}$  &   $^{+7.8\times10^{-4}}_{-6.2\times10^{-4}}$   &   ${+8.8\times10^{-3}}$   &   ${+9.6\times10^{-3}}$   &   ${+7.9\times10^{-3}}$ \\
0.4--0.6   &   $+$0.032   &   ${\pm4.7\times10^{-4}}$  &   $^{+4.4\times10^{-4}}_{-4.9\times10^{-4}}$   &   ${+6.3\times10^{-3}}$   &   ${+7.2\times10^{-3}}$   &   ${+5.5\times10^{-3}}$ \\
0.6--0.8   &   $+$0.015   &   ${\pm5.1\times10^{-4}}$  &   $^{+3.6\times10^{-4}}_{-4.5\times10^{-4}}$   &   ${+4.3\times10^{-3}}$   &   ${+5.3\times10^{-3}}$   &   ${+3.3\times10^{-3}}$ \\
0.8--1   &   $+$0.008   &   ${\pm5.5\times10^{-4}}$  &   $^{+3.8\times10^{-4}}_{-4.9\times10^{-4}}$   &   ${+3.1\times10^{-3}}$   &   ${+4.1\times10^{-3}}$   &   ${+2.1\times10^{-3}}$ \\
1--1.2   &   $+$0.002   &   ${\pm6.0\times10^{-4}}$  &   $^{+3.6\times10^{-4}}_{-4.2\times10^{-4}}$   &   ${+2.5\times10^{-3}}$   &   ${+3.6\times10^{-3}}$   &   ${+1.3\times10^{-3}}$ \\
1.2--1.4   &   $+$0.002   &   ${\pm6.5\times10^{-4}}$  &   $^{+3.3\times10^{-4}}_{-2.9\times10^{-4}}$   &   ${+2.2\times10^{-3}}$   &   ${+3.2\times10^{-3}}$   &   ${+1.2\times10^{-3}}$ \\
1.4--1.6   &   $-$0.000   &   ${\pm7.0\times10^{-4}}$  &   $^{+4.2\times10^{-4}}_{-3.6\times10^{-4}}$   &   ${+1.9\times10^{-3}}$   &   ${+2.8\times10^{-3}}$   &   ${+9.8\times10^{-4}}$ \\
\end{longtable}
\end{center}
\end{small}

%cn vs deta

\clearpage
\begin{small}
\begin{center}
\renewcommand*{\arraystretch}{1.2}
\begin{longtable}{| c | c | c | c | c | c | c |}
\hline

\hline\multicolumn{1}{|c|}{$\meanpt$ (\textrm{GeV})} & \multicolumn{1}{|c|}{$c_1\{2\}$} & \multicolumn{1}{|c|}{$\delta_{\textrm{stat}}$} & \multicolumn{1}{|c|}{$\delta_{\textrm{syst}}$} & \multicolumn{1}{|c|}{$\delta_{\textrm{nc}}$} & \multicolumn{1}{|c|}{$\delta_{\textrm{nc}}^{\textrm{ARIADNE}}$} & \multicolumn{1}{|c|}{$\delta_{\textrm{nc}}^{\textrm{LEPTO}}$} \\ \hline 
\endfirsthead

\multicolumn{7}{c}%
{ \tablename\ \thetable{} -- continued from previous page} \\
\hline
\endhead

\hline
\multicolumn{7}{c}%
{ \tablename\ \thetable{} -- continued on next page} \\
\hline
\endfoot

\endlastfoot 

0.1--0.2   &   $-$0.01   &   ${\pm8.6\times10^{-4}}$  &   $^{+2.3\times10^{-3}}_{-2.2\times10^{-3}}$   &   ${+1.4\times10^{-3}}$   &   ${+1.2\times10^{-3}}$   &   ${+1.5\times10^{-3}}$ \\
0.2--0.4   &   $-$0.02   &   ${\pm3.6\times10^{-4}}$  &   $^{+2.2\times10^{-3}}_{-2.2\times10^{-3}}$   &   ${+2.4\times10^{-3}}$   &   ${+2.5\times10^{-3}}$   &   ${+2.4\times10^{-3}}$ \\
0.4--0.6   &   $-$0.05   &   ${\pm4.1\times10^{-4}}$  &   $^{+1.9\times10^{-3}}_{-2.0\times10^{-3}}$   &   ${+4.1\times10^{-3}}$   &   ${+4.1\times10^{-3}}$   &   ${+4.1\times10^{-3}}$ \\
0.6--0.8   &   $-$0.07   &   ${\pm5.4\times10^{-4}}$  &   $^{+1.8\times10^{-3}}_{-1.8\times10^{-3}}$   &   ${+4.7\times10^{-3}}$   &   ${+4.0\times10^{-3}}$   &   ${+5.4\times10^{-3}}$ \\
0.8--1   &   $-$0.10   &   ${\pm7.4\times10^{-4}}$  &   $^{+2.2\times10^{-3}}_{-2.2\times10^{-3}}$   &   ${+4.4\times10^{-3}}$   &   ${+3.0\times10^{-3}}$   &   ${+5.9\times10^{-3}}$ \\
1--1.2   &   $-$0.13   &   ${\pm9.9\times10^{-4}}$  &   $^{+2.5\times10^{-3}}_{-2.4\times10^{-3}}$   &   ${+2.5\times10^{-3}}$   &   ${+8.4\times10^{-4}}$   &   ${+4.2\times10^{-3}}$ \\
1.2--1.4   &   $-$0.15   &   ${\pm1.3\times10^{-3}}$  &   $^{+2.9\times10^{-3}}_{-2.9\times10^{-3}}$   &   ${+1.1\times10^{-3}}$   &   ${-6.7\times10^{-4}}$   &   ${+2.8\times10^{-3}}$ \\
1.4--1.6   &   $-$0.17   &   ${\pm1.7\times10^{-3}}$  &   $^{+2.1\times10^{-3}}_{-2.0\times10^{-3}}$   &   ${-3.1\times10^{-4}}$   &   ${-2.8\times10^{-3}}$   &   ${+2.2\times10^{-3}}$ \\
\end{longtable}%
\end{center}
\end{small}

\end{document}